\documentclass[fleqn,usenatbib]{mnras}
\usepackage{amsfonts}
\usepackage{graphicx}
\usepackage{amsmath}
\usepackage{amssymb}
\usepackage{multicol}
\usepackage{bm}
\usepackage{pdflscape}
\usepackage[T1]{fontenc}
\usepackage{ae,aecompl}

\pubyear{2017}

\begin{document}

\title[Nonlinear approach to entrainment]{\textit{Nonlinear approach to the
entrainment matrix of superfluid nucleon mixture at zero temperature}}
\author[Lev B. Leinson]{Lev B. Leinson$^{1}$\thanks{%
E-mail: leinson@yandex.ru} \\
%EndAName
$^{1}$Pushkov Institute of Terrestrial Magnetism, Ionosphere and Radiowave
Propagation of the Russian Academy of Science (IZMIRAN), \\
108840 Troitsk, Moscow, Russia}
\date{Accepted 5 June 2017. Received 31 May 2017; in original form 18 March 2017}
\maketitle

\begin{abstract}
The superfluid drag effect, in hydrodynamics of pulsating neutron stars, is
conventionally described with the aid of the entrainment matrix relating the
mass currents with the velocities of superfluid flows in the system.
Equations for the entrainment matrix of a superfluid mixture of neutrons and
protons are derived with allowance for the strong dependence of the energy
gaps on the velocities of superfluid flows. The calculations are carried out
in the frame of the Fermi-liquid theory. The equations obtained are highly
nonlinear. Numerical solutions to the equations for some typical cases
demonstrate that the components of the entrainment matrix possess a highly
nonlinear dependence on the velocities of the two superflows simultaneously.
This effect, previously ignored, can greatly influence the dynamics of neutron stars.
\end{abstract}

\label{firstpage} \pagerange{\pageref{firstpage}--\pageref{lastpage}}

%%%%%%%%%%%%%%%%%%%%%%%%%%%%%%%%%%%%%%%%%%

%%%%%%%%%%%%%%%%%%%%%%%%%%%%%%%%%%%%%%%%%%

%%%%%% BODY OF PAPER %%%%%%%%%%%%%%%%%%

%%\PACS 97.60.Jd, 47.37.+q, 71.10.Ay, 04.40.Dg

%%%%%%%%%%%%%%%%%%%%%%%%%%%%%%%%%%%%%%%%%%%%%%%%%%%%%%%%%%%%%%%%%%%%

\section{Introduction}

%%%%%%%%%%%%%%%%%%%%%%%%%%%%%%%%%%%%%%%%%%%%%%%%%%%%%%%%%%%%%%%%%%%%

Pulsations of neutron stars consisting mostly of superfluid nucleons are of
a great interest today %
\citep[see][]{aks98,a98,fm98,acp02,ak01,gck14,aetal03,gyg05,spa10,pr02,gj14,son16,ap17}. As
is well known the theory of this phenomenon involves the drag effects
leading to the mutual entrainment of superfluid flows of neutrons and
protons. This effect is conventionally described with the aid of the so
called "entrainment matrix" that relates the mass current density with a
relative velocity of superfluid and normal components in the nucleon liquid.
For the simplest case of the \textsf{npe}-matter the symmetric entrainment
matrix $\rho _{\alpha \beta }$ $\left( \alpha ,\beta =\mathsf{n,p}\right) $
can be defined by the Andreev-Bashkin relation \citep{ab75}: 
\begin{eqnarray}
\mathbf{j}_{\mathsf{n}} &=&\left( \rho _{\mathsf{n}}-\rho _{\mathsf{nn}%
}-\rho _{\mathsf{np}}\right) \mathbf{v}_{\mathsf{nq}}+\rho _{\mathsf{nn}}%
\mathbf{v}_{\mathsf{n}s}+\rho _{\mathsf{np}}\mathbf{v}_{\mathsf{p}s},
\label{jn} \\
\mathbf{j}_{\mathsf{p}} &=&\left( \rho _{\mathsf{p}}-\rho _{\mathsf{pp}%
}-\rho _{\mathsf{pn}}\right) \mathbf{v}_{\mathsf{pq}}+\rho _{\mathsf{pp}}%
\mathbf{v}_{\mathsf{p}s}+\rho _{\mathsf{pn}}\mathbf{v}_{\mathsf{n}s},
\label{jp}
\end{eqnarray}%
where $\rho _{\alpha }=m_{\alpha }n_{\alpha }$. Hereafter $m_{\alpha}$ and $%
n_{\alpha }$ are the nucleon mass and number density of nucleon species $%
\alpha $, respectively; $\mathbf{j}_{\alpha }$ and $\mathbf{v}_{\alpha s}$
denote the mass current densities and the velocities of neutron and proton
superfluid components. Finally, $\mathbf{v}_{\alpha \mathsf{q}}$ represent
velocities of normal (non-superfluid) nucleons.\footnote{%
The normal component vanishes at zero temperature in a superfluid at rest.
However, in superfluid flows, the normal component appears due to the
superfluid motion.} For definiteness we consider the mixture of superfluids
in a coordinate frame, where $\mathbf{v}_{\mathsf{q}}=0$.

All the existing calculations of the entrainment matrix%
\citep[see{,
e.g.}][]{Betal96,ch06,gkh09} do not take into consideration the change
of the energy gap $\Delta ^{\left( \alpha \right) }\left( \mathbf{v}_{\alpha
s}\right) $ due to\ the motion of superfluid condensates and thus are
justified only for small oscillation amplitudes, which are restricted by the
condition $v_{\alpha s}\ll v_{cr}^{\left( \alpha \right) }$, where $%
v_{cr}^{\left( \alpha \right) }\sim \Delta _{0}^{\left( \alpha \right)
}/p_{F\alpha }$ is the critical speed of the superfluid flow at which the
superfluidity is destroyed ($\Delta _{0}^{\left( \alpha \right) }$ is the
energy gap for Bogoliubov excitations in the superfluid at rest). A
comparison of the critical velocity, which is necessary for the energy gap
collapse with the known estimates of relative velocities between the
superfluid and normal components inside oscillating neutron stars %
\citep{kg14,gk13,gyg05} clearly demonstrates that the speed of superfluid motion
can exceed the critical values, which means that the considered gap
suppression effect can be very important for the dynamics of superfluid
pulsations. Our goal is to develop an approach that allows one to evaluate
the entrainment matrix at arbitrary speed of the superflows in the \textsf{%
npe}-matter.

The suggested theory assumes that protons are paired in the spin-singlet $^1$%
S$_0$ state while the neutron pairing occurs into the spin-triplet $^3$P$_2$
state \citep{Tamagaki,Takatsuka}. We assume also that, in the \textsf{npe}
matter, the coulomb interaction of protons is completely screened by
relativistic degenerate electrons.

The paper is organized as follows. In Sec. \ref{sec:dv} the problem is
considered in the frame of the Fermi liquid theory, where the Fermi-liquid
interactions are expressed through the Landau parameters. We derive the
nonlinear equations for the entrainment matrix in a mixture of superfluid
Fermi liquids, which are valid for any velocities of the superfluid flows.
The equations obtained must be solved simultaneously with the equations for
velocity-dependent energy gaps in the superfluid flows. In Sec. \ref{sec:gap} we
solve the corresponding gap equations for protons and neutrons in the case
of zero temperature. In Sec. \ref{sec:Fi} we calculate the functions
describing the balance between the superfluid and normal components in the
superfluid flows. In Sec. \ref{sec:em} we incorporate the resulting solution
into equations for the entrainment matrix. Section \ref{sec:cur} contains
a derivation of the expression for the superfluid mass currents in the
system under consideration. In Sec. \ref{sec:lim} we discuss possible
applications of the nonlinear equations. We demonstrate the results of their
numerical solution for some interesting cases and compare them with known
results obtained in the linear approximation. The summary and conclusion are
collected in Sec. \ref{sec:conc}. The appendix contains the derivation of
the equation for the $^{3}$P$_{2}$ energy gap in the moving superfluid flow.

We use the system of units in which $c=\hbar =1$, and the Boltzmann constant 
$k_{\mathrm{B}}=1$.

%%%%%%%%%%%%%%%%%%%%%%%%%%%%%%%%%%%%%%%%%%%%%%%%%%%%%%%%%%%%%%%%%%%%

\section{Nonlinear equations for the entrainment matrix}

\label{sec:dv}

%%%%%%%%%%%%%%%%%%%%%%%%%%%%%%%%%%%%%%%%%%%%%%%%%%%%%%%%%%%%%%%%%%%%
We begin with a general expression for the mass current density $\mathbf{j}%
_{\alpha }$ in a superfluid mixture of two species of baryons. For
concreteness, we shall consider protons and neutrons, $\alpha =\mathsf{n,p}$%
, although the same theory can be applied in the case of superfluid hyperons
and nucleons in the neutron star core. Following the Fermi-liquid theory,
the mass current density can be evaluated from the same expression, which is
normally used in the case of non-superfluid matter \citep{leg65,leg75} 
\begin{equation}
\mathbf{j}_{\alpha }=\sum_{\mathbf{k,}\sigma }m_{\alpha }\frac{\partial 
\tilde{\varepsilon}_{\mathbf{k}}^{(\alpha )}}{\partial \mathbf{k}}\tilde{n}_{%
\mathbf{k}}^{(\alpha )},  \label{J}
\end{equation}%
where $\mathbf{k}$ and $\sigma =\uparrow ,\downarrow $ are the momentum of the quasiparticle and its spin projection onto the quantization axis, respectively. 
In this expression, the quasiparticle energy $\tilde{%
\varepsilon}_{\mathbf{k}}^{(\alpha )}$ depends on the distribution $\tilde{n}%
_{\mathbf{k}}^{(\alpha )}$ of all interacting particles in the system in the
presence of superfluid motion. Here and below, the tilde above a letter
indicates the values that depend on the velocity of the superfluid flow.

For a system at rest, the energy of protons and neutrons in a normal Fermi
liquid $\epsilon _{k}^{(\alpha )}$ is renormalized by interactions only.
According to the Landau theory the renormalized energy of a quasiparticle
may be written with the aid of the effective mass $m_{\alpha }^{\ast }$,
which is no longer equal to the bare mass $m_{\alpha }$ but is defined by \ 
\begin{equation}
v_{F\alpha }=\frac{d\epsilon _{k}^{(\alpha )}}{dk}=\frac{k}{m_{\alpha
}^{\ast }},  \label{vF}
\end{equation}%
where the derivative is evaluated at the appropriate Fermi surface given by
the Fermi wave vector 
\begin{equation}
p_{F\alpha }=\left( 3\pi ^{2}n_{\alpha }\right) ^{1/3}.  \label{kF}
\end{equation}%
Since the nucleon matter is highly degenerate it is convenient to write the
energy of a quasiparticle $\alpha $ as%
\begin{equation}
\epsilon ^{(\alpha )}\left( k\right) =\mu _{\alpha }+\xi _{\mathbf{k}%
}^{(\alpha )},  \label{mu}
\end{equation}%
where $\mu _{\alpha }=\epsilon ^{(\alpha )}\left( p_{F\alpha }\right) $~is
the chemical potential (the Fermi energy) of a quasiparticle species $\alpha 
$, and $\xi _{k}^{(\alpha )}\equiv \epsilon ^{(\alpha )}\left( k\right) -\mu
_{\alpha }$. Near the Fermi surface one can approximate 
\begin{equation}
\xi _{k}^{(\alpha )}=v_{F\alpha }\left( k-p_{F\alpha }\right) .  \label{epk}
\end{equation}%
In the mixture of superfluid liquids at rest, the quasiparticle energy is
renormalized as\footnote{%
This correction $\sim (\Delta /\mu)^2 $ is small and normally ignored.}%
\begin{equation}
\varepsilon _{\mathbf{k}}^{(\alpha )}=\xi _{k}^{(\alpha )}+\sum_{\mathbf{k}%
^{\prime }\sigma ^{\prime }\beta }f^{\alpha \beta }\left( \mathbf{k,k}%
^{\prime }\right) \left( n_{\mathbf{k}^{\prime }}^{(\beta )}-\Theta _{%
\mathbf{k}^{\prime }}^{(\beta )}\right) ,  \label{eps0}
\end{equation}%
where $\Theta _{\mathbf{k}}^{\left( \alpha \right) }\equiv \Theta \left(
p_{F\alpha }-\left\vert \mathbf{k}\right\vert \right) $~is the step
function, and 
\begin{equation}
n_{\mathbf{k}}^{(\alpha )}\equiv \left\langle \left\vert a_{\mathbf{k}\sigma
}^{(\alpha )\dagger }a_{\mathbf{k}\sigma }^{(\alpha )}\right\vert
\right\rangle  \label{nk}
\end{equation}%
is the distribution function of quasiparticles in the mixture of superfluid
liquids at rest. For simplicity, we consider unitary states of the superfluid nucleons 
of both species. Since the uniform motion of a unitary superfluid liquid induces 
no spin polarization we shall use the spin-averaged Fermi-liquid interactions, which are parametrized by the functions 
$f^{\alpha\beta }\left( \mathbf{k,k}^{\prime }\right) $.

The supercurrent in the system arises when the Cooper condensate is formed
by pairing of quasiparticles with momentum $(\mathbf{k+q}_{\alpha })$ and $(-%
\mathbf{k+q}_{\alpha })$. In this case, each pair acquires a nonzero
momentum $2\mathbf{q}_{\alpha }$, and the condensate moves, as a whole, with
a velocity $\mathbf{v}_{\mathsf{\alpha }s}=\mathbf{q}_{\alpha }/m_{\alpha }$
relative to non-superfluid components, if any.

The energy of quasiparticles participating in the superfluid motion is
renormalized according to 
\begin{equation}
\tilde{\varepsilon}_{\mathbf{k+q}_{\alpha }}^{(\alpha )}=\xi _{k}^{(\alpha
)}+\frac{\mathbf{kq}_{\alpha }}{m_{\alpha }^{\ast }}+\sum_{\mathbf{k}%
^{\prime }\sigma ^{\prime }\beta }f^{\alpha \beta }\left( \mathbf{k,k}%
^{\prime }\right) \left( \tilde{n}_{\mathbf{k}^{\prime }\mathbf{+q}_{\beta
}}^{(\beta )}-\Theta _{\mathbf{k}^{\prime }\mathbf{+q}_{\beta }}^{(\beta
)}\right) ,  \label{epsEq}
\end{equation}%
Hereafter we use the fact that inequalities\ $q_{\alpha }\ll p_{F\alpha }$,
or equivalently, $v_{\alpha s}\ll v_{F\alpha }$ are well fulfilled, in
superfluid Fermi liquids. Therefore we restrict ourselves to a linear in $%
q_{\alpha }/p_{F\alpha }$ terms to write $f^{\alpha \beta }\left( \mathbf{k%
\mathbf{+q}_{\alpha },k}^{\prime }\mathbf{+q}_{\beta }\right) \simeq
f^{\alpha \beta }\left( \mathbf{k,k}^{\prime }\right) $.

Combining Eqs. (\ref{eps0}) and (\ref{epsEq}) one finds that the
renormalization of the quasiparticle energy owing to the motion of
superfluids is of the form%
\begin{equation}
\delta \varepsilon _{\mathbf{k+q}_{\alpha }}^{(\alpha )}=\frac{\mathbf{kq}%
_{\alpha }}{m_{\alpha }^{\ast }}+\sum_{\mathbf{k}^{\prime }\beta \sigma
^{\prime }}f^{\alpha \beta }\left( \mathbf{k,k}^{\prime }\right) \delta n_{%
\mathbf{k}^{\prime }\mathbf{+q}_{\beta }}^{(\beta )},  \label{epsq}
\end{equation}%
where the change of the distribution functions because of the supercurrents
is given by%
\begin{equation}
\delta n_{\mathbf{k+q}_{\alpha }}^{(\alpha )}\equiv \left( \tilde{n}_{%
\mathbf{k+q}_{\alpha }}^{(\alpha )}-\Theta _{\mathbf{k+q}_{\alpha
}}^{(\alpha )}\right) -\left( n_{\mathbf{k}}^{(\alpha )}-\Theta _{\mathbf{k}%
}^{(\alpha )}\right) .  \label{dn}
\end{equation}

All existing computations of the entrainment matrix 
\citep[see{,
e.g.}][]{Betal96, ch06, gkh09} do not take into account the change in
the energy gap $\Delta ^{\left( \alpha \right) }\left( \mathbf{q}%
_{\alpha}\right)$ due to the motion of superfluid condensates and,
consequently, are justified only for small amplitudes of oscillations that
are restricted by the condition $v_{\alpha s}\ll v_{cr}^{\left( \alpha
\right) }$, where $v_{cr}^{\left( \alpha \right) }\sim \Delta _{0}^{\left(
\alpha \right) }/p_{F\alpha }$ is the critical velocity of the superfluid
flow at which the superfluidity is destroyed ($\Delta _{0}^{\left( \alpha
\right) }$ is the energy gap for Bogoliubov excitations in the superfluid at
rest).

To develop the theory, which is valid in the case of near-critical
supercurrents one needs to incorporate, into the distribution function, the
elementary excitations arising in the system due to the motion of the
superfluid condensate. To this end, it is necessary to express the
quasiparticle operators $a_{\mathbf{k+q}_{\alpha }\uparrow }^{(\alpha )}$
and $a_{\mathbf{k+q}_{\alpha }\downarrow }^{(\alpha )}$, in the distribution
function definition 
\begin{equation}
\tilde{n}_{\mathbf{k+q}_{\alpha }}^{(\alpha )}\equiv \left\langle \left\vert
a_{\mathbf{k+q}_{\alpha }\sigma }^{(\alpha )\dagger }a_{\mathbf{k+q}_{\alpha
}\sigma }^{(\alpha )}\right\vert \right\rangle ,  \label{nq}
\end{equation}%
in terms of Bogoliubov excitation operators $b_{\pm \mathbf{k+q}_{\alpha
}\sigma }^{(\alpha )}$ \citep[see{, e.g.}][]{lp80}. The corresponding
conversion has been carried out by \citet{gh05}. We shall only outline their
result, referring the reader to their original work for more details. As it
follows from Eqs. (26)-(30) of the work \citep{gh05} the distribution
function (\ref{nq}) can be recast to the form 
\begin{equation}
\tilde{n}_{\mathbf{k+q}_{\alpha }}^{(\alpha )}=\tilde{u}_{\mathbf{k}%
}^{(\alpha )2}\mathcal{F}_{\mathbf{k+q}_{\alpha }}^{(\alpha )}+\tilde{v}_{%
\mathbf{k}}^{(\alpha )2}\left( 1-\mathcal{F}_{-\mathbf{k+q}_{\alpha
}}^{(\alpha )}\right) ,  \label{nqb}
\end{equation}%
where the functions%
\begin{equation}
\mathcal{F}_{\pm \mathbf{k+q}_{\alpha }}^{(\alpha )}=\frac{1}{1+e^{\frac{1}{T%
}\mathfrak{E}_{\pm \mathbf{k+q}_{\alpha }}^{\left( \alpha \right) }}}
\label{f}
\end{equation}%
represent the occupation numbers of Bogoliubov excitations of the energy $%
\mathfrak{E}_{\pm \mathbf{k+q}_{\alpha }}^{\left( \alpha \right) }$ in the
system at the temperature $T$. The parameters $u_{\mathbf{k}}^{(\alpha )}$
and $v_{\mathbf{k}}^{(\alpha )}$ are defined as%
\begin{equation}
\tilde{u}_{\mathbf{k}}^{(\alpha )2}=\frac{1}{2}\left( 1+\frac{\xi _{\mathbf{k%
}}^{(\alpha )}}{\tilde{E}_{k}^{\left( \alpha \right) }}\right) ,  \label{uk}
\end{equation}%
and%
\begin{equation}
\tilde{v}_{\mathbf{k}}^{(\alpha )2}=\frac{1}{2}\left( 1-\frac{\xi _{\mathbf{k%
}}^{(\alpha )}}{\tilde{E}_{k}^{\left( \alpha \right) }}\right) .  \label{vk}
\end{equation}%
Hereafter%
\begin{equation}
\tilde{E}_{k}^{\left( \alpha \right) }\equiv \sqrt{\xi _{\mathbf{k}%
}^{(\alpha )2}{}+\tilde{\Delta}^{(\alpha )\,2}},  \label{Ekt}
\end{equation}%
and $\tilde{\Delta}^{(\alpha )}$ denotes the energy gap, which depends on
the superflow velocity.\footnote{%
Below it will be seen that the energy gap of each species $\alpha $ depends
on the velocities of both superfluid flows, $\mathbf{v}_{\mathsf{p}s}$ and $%
\mathbf{v}_{\mathsf{n}s}$.}

If the system is at rest, the Bogoliubov excitations are absent at zero
temperature and the functions $\mathcal{F}_{\pm \mathbf{k}}^{(\alpha )}$
vanish, thus 
\begin{equation}
n_{\mathbf{k}}^{(\alpha )}=\frac{1}{2}\left( 1-\frac{\xi _{\mathbf{k}%
}^{(\alpha )}}{E_{k}^{\left( \alpha \right) }}\right) ,  \label{nrest}
\end{equation}%
where 
\begin{equation}
E_{k}^{\left( \alpha \right) }\equiv \sqrt{\xi _{\mathbf{k}}^{(\alpha
)2}{}+\Delta _{0}^{(\alpha )\,2}}  \label{Eka}
\end{equation}%
is the energy of Bogoliubov excitations in the superfluid at rest.

In a superfluid flow, the functions $\mathcal{F}_{\pm \mathbf{k}}^{(\alpha
)} $ does not vanish even at zero temperature. To demonstrate this let us
expand the quasiparticle energy variation due to the superfluid motion up to
the first order in the small parameter $q_{\alpha }/k_{F\alpha }\ll 1$. 
\begin{equation}
\delta \varepsilon _{\mathbf{k+q}_{\alpha }}^{(\alpha )}=\sum_{\beta }\gamma
_{\alpha \beta }\left( \mathbf{v}_{\mathsf{p}s},\mathbf{v}_{\mathsf{n}%
s}\right) \mathbf{kv}_{\beta s},  \label{deps}
\end{equation}%
where $\gamma _{\alpha \beta }\left( \mathbf{v}_{\mathsf{p}s},\mathbf{v}_{%
\mathsf{n}s}\right) $ is some unknown matrix, which is taken at the Fermi
surface of particle species (at $k=p_{F\beta }$) and depends on the flow
velocities according to the relations $\mathbf{q}_{\alpha }\equiv m_{\alpha }%
\mathbf{v}_{\alpha s}$. Notice, the dependence of the matrix 
$\gamma_{\alpha \beta }$ on the flow velocities reflects the balance between 
the superfluid and normal components in the flow. This balance changes along 
with the flow velocity change and should be necessarily taken into account 
(see Sec. \ref{sec:Fi}).

In the case of spin-singlet isotropic pairing of
protons the form (\ref{deps}) is quite general because there are only two
vectors $\mathbf{k}$ and $\mathbf{v}_{\mathsf{p}s}$ to form a scalar $\delta
\varepsilon _{\mathbf{k+q}_{\alpha }}^{(\alpha )}$. The same form seems
justified also in the case of anisotropic triplet pairing of neutrons if one
assumes that the gap function is axial symmetric and that the preferred
direction for the principal-axis of the gap matrix is specified by the
direction of superflow motion, i.e. $Oz\parallel \mathbf{v}_{\mathsf{n}s}$.
At least at zero temperature this choice of the principal axis seems
reasonable because the direction of the flow is the only preferred direction
which exists in the uniform system in the absence of external fields (see appendix).

Neglecting terms of the order of 
$(\Delta ^{\left( \alpha \right)}/\mu_{\alpha })^2$
one can write $\varepsilon _{\mathbf{k}}^{(\alpha )}\simeq \xi
_{k}^{(\alpha )}$. Than the quasiparticle energy (\ref{epsEq}) can be
expanded as%
\begin{equation}
\tilde{\varepsilon}_{\mathbf{k+q}_{\alpha }}^{(\alpha )}=\xi _{k}^{(\alpha
)}+\sum_{\beta }\gamma _{\alpha \beta }\left( \mathbf{v}_{\mathsf{p}s},%
\mathbf{v}_{\mathsf{n}s}\right) \mathbf{kv}_{\beta s},  \label{gam}
\end{equation}%
Equations for the unknown components of the matrix $\gamma _{\alpha \beta }$
can be obtained by combining of Eqs. (\ref{epsq}) and (\ref{deps}). To this
end one needs to calculate the change of the distribution functions owing to
the supercurrents as is given in Eq. (\ref{dn}). We proceed to this
calculation with the aid of Eqs. (\ref{nqb}) - (\ref{Ekt}).

Inserting Eqs. (\ref{nqb}), (\ref{uk}), (\ref{vk}), and (\ref{nrest}) into
Eq. (\ref{dn}) we get 
\begin{align}
\delta n_{\mathbf{k+q}_{\alpha }}^{(\alpha )}& =\frac{\xi _{\mathbf{k}%
}^{(\alpha )}}{2E_{k}^{\left( \alpha \right) }}-\frac{\xi _{\mathbf{k}%
}^{(\alpha )}}{2\tilde{E}_{k}^{\left( \alpha \right) }}\left( 1-\mathcal{F}_{%
\mathbf{k+q}_{\alpha }}^{(\alpha )}-\mathcal{F}_{-\mathbf{k+q}_{\alpha
}}^{(\alpha )}\right)  \notag \\
& +\frac{1}{2}\left( \mathcal{F}_{\mathbf{k+q}_{\alpha }}^{(\alpha )}-%
\mathcal{F}_{-\mathbf{k+q}_{\alpha }}^{(\alpha )}\right) +\mathbf{q}_{\alpha
}\mathbf{\hat{k}\,}\delta \left( k-p_{F\alpha }\right) ,  \label{deln}
\end{align}%
where $\mathbf{\hat{k}\,=k/}k$ denotes unit vector in the $\mathbf{k\,}$\
direction.

In Eq. (\ref{deln}) the last term implements the fact that the relations $%
q_{\alpha }\ll p_{F\alpha }$, or equivalently, $v_{\alpha s}\ll v_{F\alpha }$
are well fulfilled, in superfluid Fermi liquids, and one can therefore keep
only the terms linear in $\mathbf{q}_{\alpha }$, thus obtaining $\Theta _{%
\mathbf{k+q}}^{\left( \alpha \right) }-\Theta _{k}^{\left( \alpha \right)
}\simeq -\mathbf{q}_{\alpha }\mathbf{\hat{k}\,}\delta \left( k-p_{F\alpha
}\right) $.

General form for the energy of Bogoliubov excitations in the mixture of
neutrons and protons with superfluid currents was found in \citet{gh05} by a
minimization of the thermodynamic potential of the system. From Eqs. (29),
(33), and (34) of that work one finds

\begin{equation}
\mathfrak{E}_{\pm \mathbf{k+q}_{\alpha }}^{\left( \alpha \right) }=\tilde{E}%
_{k}^{\left( \alpha \right) }\pm \sum_{\beta }\gamma _{\alpha \beta }\mathbf{%
kv}_{\beta s}.  \label{Eq}
\end{equation}%
It is easily to see that this energy becomes negative provided%
\begin{equation}
\tilde{E}_{k}^{\left( \alpha \right) }\pm \sum_{\beta }\gamma _{\alpha \beta
}\mathbf{kv}_{\beta s}\leq 0.  \label{cond}
\end{equation}%
The system becomes unstable when the flow velocity exceeds some critical
value. From Eq. (\ref{cond}) it follows that if $\tilde{E}_{k}^{\left(
\alpha \right) }-\sum_{\beta }\gamma _{\alpha \beta }\mathbf{kv}_{\beta
s}\leq 0$ for $\mathbf{kv}_{\beta s}>0$ then $\tilde{E}_{k}^{\left( \alpha
\right) }+\sum_{\beta }\gamma _{\alpha \beta }\mathbf{kv}_{\beta s}\leq 0$
for $\mathbf{kv}_{\beta s}<0$. When the energy of the Bogoliubov excitations
vanishes, they are spontaneously created in pairs and accumulate in a
superfluid system at zero temperature. As a result the distribution function
for Bogoliubov excitations in the flow takes the form 
\begin{equation}
\mathcal{F}_{\pm \mathbf{k+q}_{\alpha }}^{(\alpha )}=\Theta \left( \mp
\sum_{\beta }\gamma _{\alpha \beta }\mathbf{kv}_{\beta s}-\tilde{E}%
_{k}^{\left( \alpha \right) }\right) ,  \label{Fq}
\end{equation}%
where $\Theta \left( x\right) $ is the Heaviside step-function.

As the velocities of superfluid flows are small in a scale of the Fermi
velocities the arguments of the functions $f^{\alpha \beta }\left( \mathbf{%
k,k}^{\prime }\right) $ can be approximately put equal to their values at
the corresponding Fermi surfaces. This allows one to write the interaction
function in the form of expansion in Legendre polynomials parametrized by
the Landau parameters $f_{l}^{\alpha \beta }$: 
\begin{equation}
f^{\alpha \beta }\left( \mathbf{k,k}^{\prime }\right) =\sum_{l}f_{l}^{\alpha
\beta }P_{l}\left( \cos \theta \right) ,~\ \ \cos \theta \equiv \mathbf{\hat{%
k}\hat{k}}^{\prime }.  \label{FLint}
\end{equation}%
For a nucleon matter the Landau parameters were calculated for various
mean-field models in a series of papers %
\citep[see][]{matsui81,hm87,cgl01,cgl02,cgl03}, where it is shown that only
first two spin-averaged Landau parameters are non-zero,\footnote{%
The Landau parameters with $l\geq 2$ are actually non-vanishing in nucleon
matter, although they do not appear in simplified mean-field models.} i.e., 
$f_{l}^{\alpha \beta }=0$ at $l\geq 2$. The same was found for
nucleon-hyperon matter \citep{gkh09}. In view of this observation, one can
employ only the parameters $f_{0}^{\alpha \beta }$ and $f_{1}^{\alpha \beta
} $.

After summation over $\mathbf{k}^{\prime }$ in Eq. (\ref{epsq}) the terms
proportional to $f_{0}^{\alpha \beta }$ mutually cancel with the identical
correction to the chemical potential.\footnote{%
In the system at rest, the quasiparticle energy is defined as $\xi
_{k}^{(\alpha )}\equiv \epsilon ^{(\alpha )}\left( k\right) -\mu _{\alpha }$%
, therefore, simultaneously with the energy correction $\delta \epsilon
^{(\alpha )}$ one has to take into account the correction to the chemical
potential $\delta \mu _{\alpha }$, caused by the same interactions. Since
the $f_{0}^{\alpha \beta }$ contribution reduces to a constant mean field,
it will give the identical additive to the energy and to the chemical
potential, which mutually cancel in the expression for $\xi _{k}^{(\alpha )}$%
.} Thus, only $f_{1}^{\alpha \beta }$ are to be taken into account. This
contribution into the interaction (\ref{FLint}) is odd with respect to the
replacement $\mathbf{\hat{k}}^{\prime }\rightarrow -\mathbf{\hat{k}}^{\prime
}$. Therefore, in Eq. (\ref{deln}) only odd terms, which are shown in the
second line, will survive after substitution into Eq. (\ref{epsq}) and
subsequent integration over the solid angle $\mathbf{\hat{k}}^{\prime }$.
Making use of this fact one can derive equations for the unknown matrix $%
\gamma _{\alpha \beta }\left( \mathbf{v}_{\mathsf{p}s},\mathbf{v}_{\mathsf{n}%
s}\right) $.

To this end we define the standard dimensionless Landau parameters%
\begin{equation}
F_{1}^{\alpha \beta }\equiv f_{1}^{\alpha \beta }\sqrt{N_{0\alpha }N_{0\beta
}},  \label{Fab}
\end{equation}%
where $N_{0\alpha }=m_{\alpha }^{\ast }p_{F\alpha }/\pi ^{2}$, and introduce
two auxiliary functions $\left( \alpha =\mathsf{p},\mathsf{n}\right) $: 
\begin{equation}
\mathbf{V}_{\alpha }\left( \mathbf{v}_{\mathsf{p}s},\mathbf{v}_{\mathsf{n}%
s}\right) \equiv \sum_{\beta }\gamma _{\alpha \beta }\left( \mathbf{v}_{%
\mathsf{p}s},\mathbf{v}_{\mathsf{n}s}\right) \mathbf{v}_{\beta s}.  \label{V}
\end{equation}%
Substituting formulas (\ref{deps}) and (\ref{deln}) into Eq. (\ref{epsq}) we
get%
\begin{align}
\sum_{\beta }\gamma _{\alpha \beta }\mathbf{kv}_{\beta s}& =\frac{m_{\alpha }%
}{m_{\alpha }^{\ast }}\mathbf{kv}_{\alpha s}  \notag \\
& +\frac{1}{\sqrt{N_{0\alpha }N_{0\beta }}}\sum_{\mathbf{k}^{\prime }
\sigma^{\prime }\beta}F_{1}^{\alpha \beta }\mathbf{\hat{k}\hat{k}}^{\prime }
\left[ \mathcal{F}_{%
\mathbf{k^{\prime }+q}_{\beta }}^{(\beta )}-\mathcal{F}_
{-\mathbf{k^{\prime }+q}_{\beta}}^{(\beta )}\right.  \notag \\
& \left. +2m_{\beta }\mathbf{\hat{k}}^{\prime }\mathbf{v}_{\beta s}\mathbf{\,%
}\delta \left( k^{\prime }-k_{F\beta }\right) \right] ,  \label{geq}
\end{align}%
where 
\begin{equation}
\mathcal{F}_{\pm \mathbf{k+q}_{\alpha }}^{(\alpha )}=\Theta \left( \pm 
\mathbf{kV}_{\alpha }-\tilde{E}_{k}^{\left( \alpha \right) }\right)
\label{Fkq}
\end{equation}

In a standard way the summation over $\mathbf{k}^{\prime }$ can be converted
into the integral%
\begin{equation}
\sum_{\mathbf{k}\sigma^{\prime }}\equiv \int \frac{2d^{3}k}{\left( 2\pi \right) ^{3}}\cdot
\cdot \cdot \simeq \frac{p_{F_{\alpha }}m_{\alpha }^{\ast }}{\pi ^{2}}%
\int_{-\infty }^{\infty }d\xi _{k}\int \frac{d\mathbf{\hat{k}}}{4\pi }\cdot \cdot
\cdot .  \label{sigP}
\end{equation}%
The integral with respect to the angles on the right-hand side of Eq. (\ref%
{geq}) can be carried out using the addition theorem, which is valid for the
Legendre polynomials:%
\begin{equation}
\cos \theta _{\mathbf{kk}^{\prime }}=\cos \theta _{\mathbf{kV}}\cos \theta _{%
\mathbf{k}^{\prime }\mathbf{V}}+\sin \theta _{\mathbf{kV}}\sin \theta _{%
\mathbf{k}^{\prime }\mathbf{V}}\cos \left( \phi -\phi ^{\prime }\right) .
\label{leg}
\end{equation}%
We get%
\begin{align}
\sum_{\beta }\gamma _{\alpha \beta }\mathbf{\hat{k}v}_{\beta s}& =\frac{%
m_{\alpha }}{m_{\alpha }^{\ast }}\mathbf{\hat{k}v}_{\alpha s}  \notag \\
& +\sum_{\beta }\frac{1}{3}F_{1}^{\alpha \beta }\left( \frac{p_{F\beta }}{%
p_{F\alpha }}\right) ^{3/2}\frac{m_{\beta }}{\sqrt{m_{\alpha }^{\ast
}m_{\beta }^{\ast }}}\mathbf{\hat{k}v}_{\beta s}  \notag \\
& -\sum_{\beta }\frac{1}{3}F_{1}^{\alpha \beta }\left( \frac{p_{F\beta }}{%
p_{F\alpha }}\right) ^{3/2}\sqrt{\frac{m_{\beta }^{\ast }}{m_{\alpha }^{\ast
}}}\mathbf{\hat{k}V}_{\beta }\Phi _{\beta },  \label{gamEq}
\end{align}%
where the function $\Phi _{\alpha }\left( \tilde{\Delta}^{(\alpha
)},V_{\alpha }\right) $ is given by the integral 
\begin{equation}
\Phi _{\alpha }\equiv -\frac{3}{N_{0\alpha }}\frac{1}{p_{F\alpha }V_{\alpha }%
}\int \frac{d^{3}k}{8\pi ^{3}}\left( \mathcal{F}_{\mathbf{k+q}_{\alpha
}}^{(\alpha )}-\mathcal{F}_{-\mathbf{k+q}_{\alpha }}^{(\alpha )}\right) \cos
\theta _{\mathbf{kV}_{\alpha }},  \label{Fi}
\end{equation}%
and the distribution function for Bogoliubov excitations is defined in Eq. (%
\ref{Fkq}).

To proceed we write $\mathbf{\hat{k}V}_{\alpha }$ in Eq. (\ref{gamEq})
in the form $\mathbf{\hat{k}V}_{\alpha }=\sum_{\beta }\gamma _{\alpha \beta }%
\mathbf{\hat{k}v}_{\beta s}$ and equate terms with the same $\mathbf{\hat{k}v%
}_{\alpha s}$ in the left- and right-hand sides thus obtaining the set of
equations for $\gamma _{\alpha \beta }$:%
\begin{align}
& \left( 1+\frac{1}{3}F_{1}^{\alpha \alpha }\Phi _{\alpha }\right) \gamma
_{\alpha \alpha }+\frac{1}{3}F_{1}^{\alpha \beta }\left( \frac{p_{F\beta }}{%
p_{F\alpha }}\right) ^{3/2}\sqrt{\frac{m_{\beta }^{\ast }}{m_{\alpha }^{\ast
}}}\,\Phi _{\beta }\gamma _{\beta \alpha }  \notag \\
& =\left( 1+\frac{1}{3}F_{1}^{\alpha \alpha }\right) \frac{m_{\alpha }}{%
m_{\alpha }^{\ast }},  \label{aa}
\end{align}%
\begin{align}
& \left( 1+\frac{1}{3}F_{1}^{\alpha \alpha }\Phi _{\alpha }\right) \gamma
_{\alpha \beta }+\frac{1}{3}F_{1}^{\alpha \beta }\left( \frac{p_{F\beta }}{%
p_{F\alpha }}\right) ^{3/2}\sqrt{\frac{m_{\beta }^{\ast }}{m_{\alpha }^{\ast
}}}\Phi _{\beta }\gamma _{\beta \beta }  \notag \\
& =\frac{1}{3}F_{1}^{\alpha \beta }\left( \frac{p_{F\beta }}{p_{F\alpha }}%
\right) ^{3/2}\frac{m_{\beta }}{\sqrt{m_{\alpha }^{\ast }m_{\beta }^{\ast }}}%
.  \label{ab}
\end{align}%
Here $\alpha \neq \beta $ so that if $\alpha =\mathsf{n}$ then $\beta =%
\mathsf{p}$ and vice-versa. This formula must be supplemented by an
expression for the effective mass \citep{S73,Betal96}: 
\begin{equation}
{\frac{m_{\mathsf{\alpha }}^{\ast }}{m_{\mathsf{\alpha }}}}\,\,=1+\frac{%
N_{0\alpha }}{3}\left[ f_{1}^{\mathsf{\alpha \alpha }}+\frac{m_{\beta }}{%
m_{\alpha }}\left( \frac{p_{F\beta }}{p_{F\alpha }}\right) ^{2}f_{1}^{%
\mathsf{\alpha \beta }}\right] .  \label{mef}
\end{equation}%
It is convenient to recast this relation in terms of the dimensionless
Landau parameters:%
\begin{equation}
{\frac{m_{\mathsf{\alpha }}^{\ast }}{m_{\mathsf{\alpha }}}}\,\,=\frac{\left(
3+F_{1}^{\alpha \alpha }\right) \,\left( 3+\,F_{1}^{\beta \beta }\right)
-\,F_{1}^{\alpha \beta }F_{1}^{\beta \alpha }}{3\left( 3+\,F_{1}^{\beta
\beta }-F_{1}^{\alpha \beta }p_{F\beta }/p_{F\alpha }\right) },~\ \ \ \beta
\neq \alpha  \label{mefL}
\end{equation}

The Eqs. (\ref{aa}), (\ref{ab}) should be solved simultaneously with the gap
equations in order to consistently take into account the gap dependence on
the flow velocities. Below we consider these equations.

It is necessary to stress that Eqs. (\ref{aa}), (\ref{ab}) represent a set
of nonlinear equations, where the integrals (\ref{Fi}) depend on the unknown
functions $\gamma _{\alpha \beta }$ and the flow velocities $\mathbf{v}%
_{\alpha s}$ according to Eqs. (\ref{Fkq}) and (\ref{V}).

%%%%%%%%%%%%%%%%%%%%%%%%%%%%%%%%%%%%%%%%%%%%%%%%%%%%%%%%%%%%%%%%%%%%

\section{Gap equations}

\label{sec:gap}

\subsection{$^{1}$S$_{0}$ energy gap in a superfluid flow of protons}

%%%%%%%%%%%%%%%%%%%%%%%%%%%%%%%%%%%%%%%%%%%%%%%%%%%%%%%%%%%%%%%%%%%%

In the case of spin-singlet pairing the equation for dependence of the
energy gap on the superflow velocity in a one-component superfluid is well
known and repeatedly discussed in the literature 
\citep[see{, e.g.}][]{bardeen62,alex03,gk13}. The $^{1}$S$_{0}$ gap equation for a
mixture of two interacting superfluids can be obtained in the same manner as
the equation for a one-component superfluid. To take into account the
interaction effects, it is only necessary to substitute the distribution
functions for the Bogoliubov excitations with momentum $\pm \mathbf{k+q}_{%
\mathsf{p}}$, as indicated in the equation (\ref{Fkq}). Omitting the
intermediate steps, we can write the gap equation in the form 
\begin{equation}
\frac{p_{F\mathsf{p}}m_{\mathsf{p}}^{\ast }}{\pi ^{2}}\ln \frac{\Delta
_{0}^{\left( \mathsf{p}\right) }}{\tilde{\Delta}^{(\mathsf{p})}}=\sum_{k}%
\frac{\mathcal{F}_{\mathbf{k+q}_{\mathsf{p}}}^{(\mathsf{p})}+\mathcal{F}_{-%
\mathbf{k+q}_{\mathsf{p}}}^{(\mathsf{p})}}{\sqrt{\xi _{k}{}^{2}+\tilde{\Delta%
}^{(\mathsf{p})\,2}}}.  \label{gapeq}
\end{equation}%
As in above, $\Delta _{0}^{\left( \mathsf{p}\right) }$ denotes the energy
gap in the superfluid system at rest, that is at $\mathbf{q}=0$, while $%
\tilde{\Delta}^{(\mathsf{p})\,}$, as before, represents the energy gap in
the presence of superfluid flows. %%%%%%%%%%%%
%Inspecting the gap equation
%along with Eqs. (\ref{Fkq}) and (\ref{V}) one can easily see that, in the
%mixture of two interacting superfluids, the energy gap of each species
%depends on the velocities of the two superfluid flows.
%%%%%%%%%%%%

The integral over momenta in Eq. (\ref{gapeq}) may be carried out
analytically. Then we obtain this equation in the form%
\begin{gather}
\left[ \ln \Lambda _{\mathsf{p}}+\ln \left( 1+\sqrt{1-\frac{\eta _{\mathsf{p}%
}^{2}}{\Lambda _{\mathsf{p}}^{2}}}\right) -\sqrt{1-\frac{\eta _{\mathsf{p}%
}^{2}}{\Lambda _{\mathsf{p}}^{2}}}\right] \Theta \left( 1-\frac{\eta _{%
\mathsf{p}}}{\Lambda _{\mathsf{p}}}\right)  \notag \\
+\Theta \left( \frac{\eta _{\mathsf{p}}}{\Lambda _{\mathsf{p}}}-1\right) \ln
\eta _{\mathsf{p}}=0.  \label{gap0}
\end{gather}%
Here and below we use the notations%
\begin{equation}
\eta _{\alpha }=\frac{\tilde{\Delta}^{(\alpha )\,}}{\Delta _{0}^{\left(
\alpha \right) }},~\ \ \Lambda _{\alpha }=\frac{p_{F\alpha }V_{\alpha }}{%
\Delta _{0}^{\left( \alpha \right) }}.  \label{eta}
\end{equation}%
The physical meaning of the introduced dimensionless parameters is quite
understandable: $\eta _{\alpha }$ is the ratio of the gap amplitude of the
species $\alpha =\mathsf{p,n}$ in the mixture of moving superfluids to the
energy gap in the same superfluid at rest.
\begin{figure}
\includegraphics[width=\columnwidth]{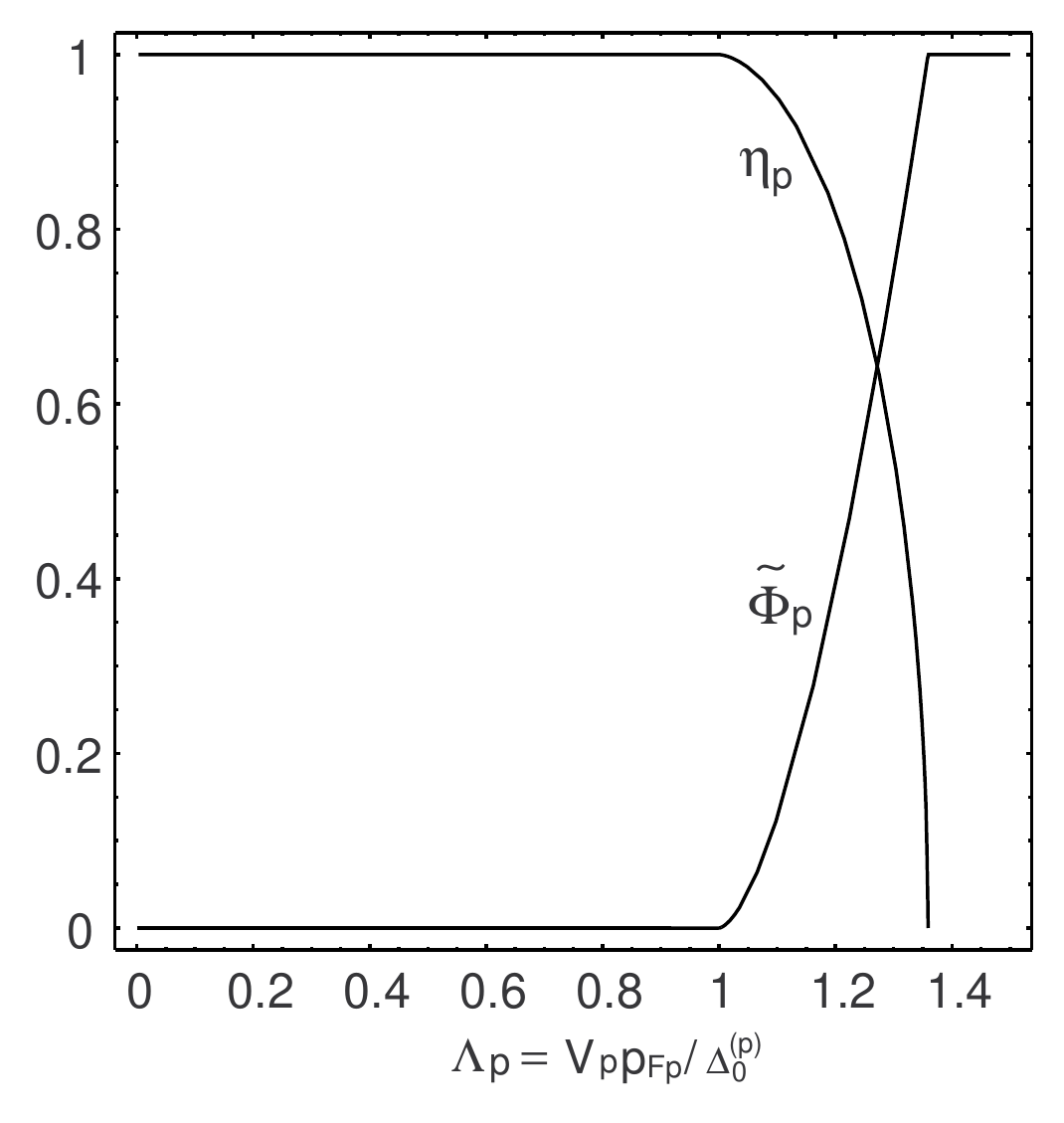}
\caption{Ratio of the energy gap in moving superfluid flows to the energy
gap in the same superfluid mixture at rest $\protect\eta _{\protect\mathsf{p} }=%
\tilde{\Delta}^{\protect\mathsf{p} }/\Delta _{0}^{\protect\mathsf{p} }$. The
function $\Tilde{\Phi}_{\protect\mathsf{p}}$ 
demonstrates an increase of the fraction of broken
pairs along with increase of the effective flow velocity $V_{\protect\mathsf{p} %
} $ above the critical value $v_{\protect\mathsf{p} 1}^{cr}=\Delta _{0}^{\left( 
\protect\mathsf{p} \right) }/p_{F\protect\mathsf{p} }$. Both the curves are
calculated for zero temperature and are shown as a function of the
dimensionless effective flow velocity $\Lambda _{\protect\mathsf{p} }=V_{\protect%
\mathsf{p} }/v_{\protect\mathsf{p} 1}^{cr}$. }
\label{fig:figeta}
\end{figure}
The parameter $\Lambda _{\alpha }$
is the effective flow velocity, 
\begin{equation}
V_{\alpha }=\left\vert \sum_{\beta }\gamma _{\alpha \beta }\mathbf{v}_{\beta
s}\right\vert ,  \label{absV}
\end{equation}%
in units of the characteristic velocity $\Delta _{0}^{\left( \alpha \right)
}/p_{F\alpha }$. It is necessary to remind that the effective velocity $V_{%
\mathsf{p}}$ in Eq. (\ref{eta}) is defined by Eq. (\ref{absV}). Thus, in a
mixture of two interacting superfluid flows, the energy gap of each species
depends on the two velocities simultaneously.

Solution to Eq. (\ref{gap0}) is depicted in Fig. \ref{fig:figeta}, where the
curve $\eta _{\mathsf{p}}$ is shown against the effective velocity $V_{%
\mathsf{p}}$ in units of $\Delta _{0}^{\left( \mathsf{p}\right) }/p_{F%
\mathsf{p}}$. 

One can see that at zero temperature there are two critical velocities. The
first one, 
\begin{equation}
\upsilon _{\mathsf{p}1}^{cr}=\frac{\Delta _{0}^{\left( \mathsf{p}\right) }}{%
p_{F\mathsf{p}}}  \label{Vc1}
\end{equation}%
represents the flow velocity at which the Bogoliubov excitations appear in
the superfluid system, and the energy gap begins to decrease.\footnote{%
At finite temperatures, $0<T<T_{c}$, the first critical velocity is smoothed
because of the temperature breaking of the Cooper pairs.} The number of
broken Cooper pairs increases rapidly when the flow velocity builds up.
Finally, all pairs are destroyed when the effective velocity reaches the
second critical value 
\begin{equation}
\upsilon _{\mathsf{p}2}^{cr}=\frac{e}{2}\frac{\Delta _{0}^{\left( \mathsf{p}%
\right) }}{p_{F\mathsf{p}}}.  \label{Vc2}
\end{equation}%
This is the well known critical velocity at which the energy gap $\tilde{%
\Delta}^{\left( \mathsf{p}\right) }$ vanishes when the Fermi-liquid
interactions are ignored. \citep[see{, e.g.}][]{alex03,gk13}.

%%%%%%%%%%%%%%%%%%%%%%%%%%%%%%%%%%%%%%%%%%%%%%%%%%%%%%%%%%%%%%%%%%%%%%%%%%%%%%%%%%%%

\subsection{$^{3}$P$_{2}$ energy gap in a superfluid flow}

%%%%%%%%%%%%%%%%%%%%%%%%%%%%%%%%%%%%%%%%%%%%%%%%%%%%%%%%%%%%%%%%%%%%%%%%%%%%%%%%%%%%

We restrict our consideration to a non rotating neutron star
and consider the unitary $^{3}$P$_{2}$ states of the gap matrix.\footnote{%
The unitarity condition implies that the superfluid state under
consideration retains time reversal symmetry and does not have, for example,
spin polarization.}  In this case the equation for the amplitude of the energy gap
$\tilde{\Delta}^{\left( \mathsf{n}\right) }$ in the superfluid neutron flow
at zero temperature can be written as (see appendix) 
\begin{equation}
\frac{p_{F}m^{\ast }}{\pi ^{2}}\ln \frac{\Delta _{0}^{\left( \mathsf{n}%
\right) }}{\tilde{\Delta}^{\left( \mathsf{n}\right) }}=\sum_{\mathbf{k}}%
\frac{\mathbf{\bar{b}}^{2}\left( \mathbf{\hat{k}}\right) }{\tilde{E}_{%
\mathbf{k}}}\left( \mathcal{F}^{\left( \mathsf{n}%
\right) }_{\mathbf{k+q}}+\mathcal{F}^{\left( \mathsf{n}%
\right) }_{-\mathbf{k+q}%
}\right) ,  \label{trgap}
\end{equation}%
where $\Delta _{0}^{\left( \mathsf{n}\right) }$ is the energy gap in the
superfluid at rest, and the distribution functions for the Bogoliubov
excitations with momentum $\pm \mathbf{k+q}_{\mathsf{n}}$, are as indicated
in Eq. (\ref{Fkq}): 
\begin{equation}
\mathcal{F}_{\pm \mathbf{k+q}_{\mathsf{n}}}^{(\mathsf{n})}=\Theta \left( \pm 
\mathbf{kV}_{\mathsf{n}}-\tilde{E}_{k}^{\left( \mathsf{n}\right) }\right)
\label{Flqn}
\end{equation}%
with%
\begin{equation}
\tilde{E}_{\mathbf{k}}^{\left( \mathsf{n}\right) }=\sqrt{\xi _{k}^{2}+\tilde{%
\Delta}^{\left( \mathsf{n}\right) 2}\,\mathbf{\bar{b}}^{2}\left( \mathbf{%
\hat{k}}\right) }.  \label{Ekn}
\end{equation}%
Here $\mathbf{\bar{b}}\left( \mathbf{\hat{k}}\right) $ is some real vector
in spin space which is normalized by the condition (\ref{Norm}). It
describes the angular dependence of the anisotropic energy gap which is
given by $D_{\mathbf{\hat{k}}}=\tilde{\Delta}^{\left( \mathsf{n}\right)
}\,\left\vert \mathbf{\bar{b}}\left( \mathbf{\hat{k}}\right) \right\vert $.
A specific form of this vector depends on the phase state of the Cooper
condensate. We restrict our consideration to the $^{3}$P$_{2}$ pairing with
a maximal projection $\left\vert M\right\vert =2$ of the total angular
momentum onto the quantization axis which is directed along the superfluid
flow motion (Arguments in favour of such a model, see appendix). In this case
one has 
\begin{equation}
\mathbf{\bar{b}}^{2}\left( \mathbf{\hat{k}}\right) =\frac{3}{2}\sin
^{2}\theta ,  \label{bk}
\end{equation}%
where $\theta $ is the angle between the quasiparticle momentum and
direction of the superflow motion. It should be noted that, by virtue of Eq.
(\ref{Norm}), the amplitudes $\Delta _{0}$ and $\tilde{\Delta}$ are chosen
as to represent the energy gap averaged over the solid angle. Defined in
this way, the average energy gap furnishes an overall measure of the pairing
correction to the ground-state energy in the preferred state.\footnote{%
It is necessary to notice that a definition of the gap amplitude is ambiguous
in the literature. For example, in the case of $\left\vert M\right\vert =2$,
our gap amplitude is $\sqrt{2/3}$ times larger than the gap amplitude in
Ref. \cite{YKL}.
Ratio $\Delta _{0}^{\left( \left\vert M\right\vert =2\right)
}/T_{c}=1.\,\allowbreak 6573$ differ in the same proportion from those
reported in \cite{ykgh01}.}

\begin{figure}
\includegraphics[width=\columnwidth]{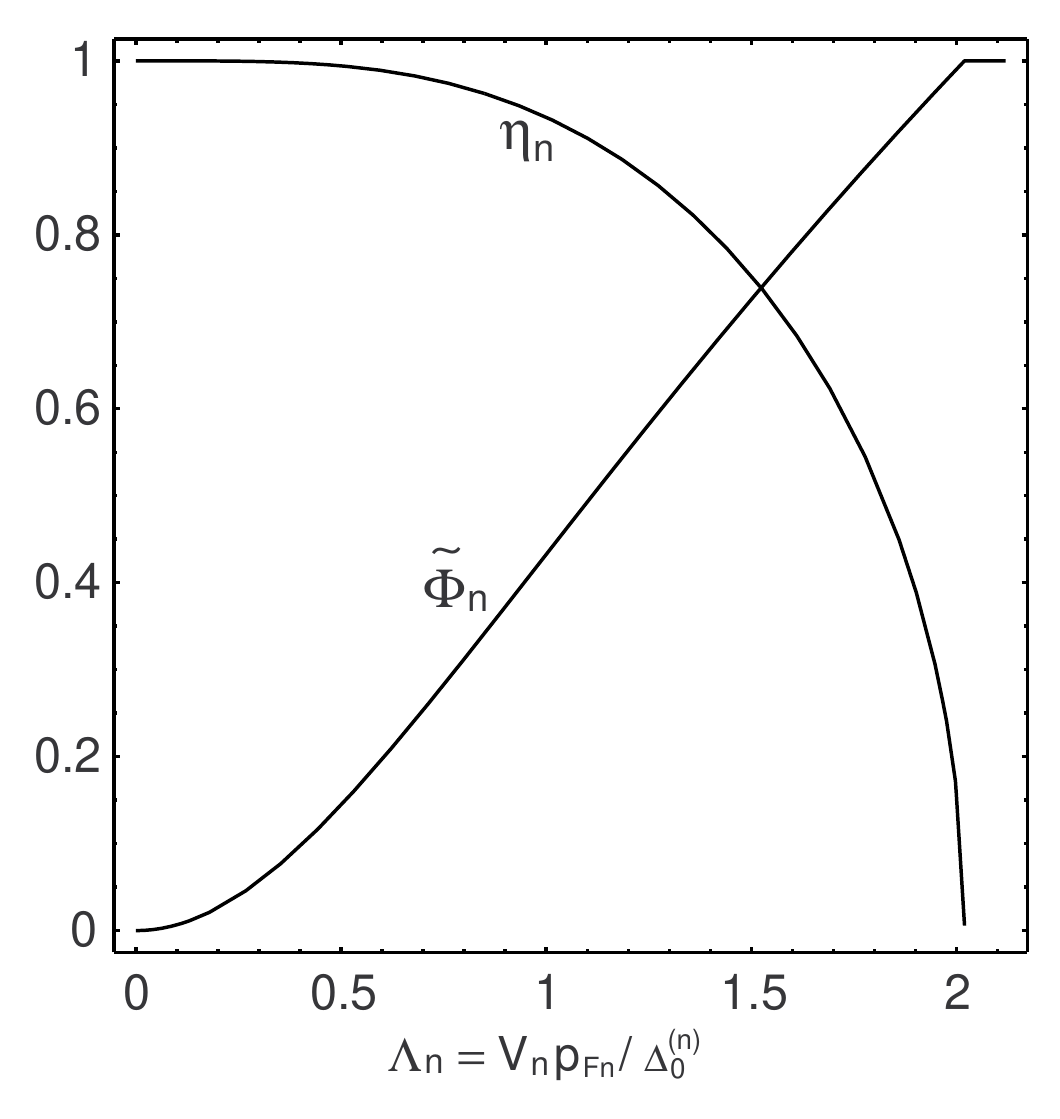}
\caption{Ratio of the energy gap in moving superfluid flows to the energy
gap in the same superfluid mixture at rest $\protect\eta _{\mathsf{n}}=%
\tilde{\Delta}^{\left( \mathsf{n}\right) }/\Delta _{0}^{\left( n\right) }$.
The function $\Tilde{\Phi}_{\mathsf{n}}$ demonstrates an increase of the
fraction of broken pairs along with increase of the effective flow velocity $%
V_{\mathsf{n}}$. Both the curves are
calculated for zero temperature and are shown as a function of the
dimensionless effective flow velocity $\Lambda _{\mathsf{n}}=V_{\mathsf{n}%
}p_{F\mathsf{n}}/\Delta _{0}^{\left( \mathsf{n}\right) }$. }
\label{fig:figeta2}
\end{figure}

The integral over momenta in Eq. (\ref{trgap}) may be carried out
analytically. Then we obtain this equation in the form%
\begin{equation}
\ln \left( \frac{2}{3}\Lambda _{\mathsf{n}}^{2}\right) -\left( 1+\frac{3}{2}%
\frac{\eta _{\mathsf{n}}^{2}}{\Lambda _{\mathsf{n}}^{2}}\right) ^{-1}+\ln
\left( 1+\frac{3}{2}\frac{\eta _{n}^{2}}{\Lambda _{n}^{2}}\right) =0
\label{gap2}
\end{equation}%
Solution to Eq. (\ref{gap2}) is depicted in Fig. \ref{fig:figeta2}, where
the curve $\eta _{\mathsf{n}}$ is shown against the effective velocity $V_{%
\mathsf{n}}$ in units of $\Delta _{0}^{\left( \mathsf{n}\right) }/p_{F%
\mathsf{n}}$. 

Important to notice that the effective velocity $\mathbf{V}_{\alpha }$ is
the auxiliary function, which is defined in Eq. (\ref{V}). It coincides with
a real velocity of the superfluid flow only in the absence of Fermi-liquid
interactions, when $\gamma _{\alpha \alpha }=1$ and $\gamma _{\alpha \beta
}=0$ for $\alpha \neq \beta $. The calculation of the entrainment matrix
dependence on the realistic velocities of superfluid flows with allowance
for Fermi-liquid interactions represents a complicated nonlinear problem. 

\section{Functions $\Phi _\alpha$}

\label{sec:Fi}

For the proton component of the superfluid mixture the integral in Eq. (\ref%
{Fi}), can be easily calculated to get%
\begin{equation}
\Phi _{\mathsf{p}}\left( \Lambda _{\mathsf{p}},\eta _{\mathsf{p}}\right)
=\left( 1-\frac{\eta _{\mathsf{p}}^{2}}{\Lambda _{\mathsf{p}}^{2}}\right)
^{3/2}\Theta \left( \Lambda _{\mathsf{p}}-\eta _{\mathsf{p}}\right) .
\label{fia}
\end{equation}%
The curve $\eta _{\mathsf{p}}\left( \Lambda _{\mathsf{p}}\right) $ on the
Fig. \ref{fig:figeta} clearly demonstrates that the condition $\Lambda _{%
\mathsf{p}}>\eta _{\mathsf{p}}$ is equivalent to the condition $\Lambda _{%
\mathsf{p}}>1$. Therefore one may replace $\Theta \left( \Lambda _{\mathsf{p}%
}-\eta _{\mathsf{p}}\right) \rightarrow \Theta \left( \Lambda _{\mathsf{p}%
}-1\right) $ in Eq. (\ref{fia}).

The simultaneous solving of Eqs. (\ref{aa}), (\ref{ab}) with the gap
equation ensures that the value of the energy gap is taken along the curve $%
\eta _{\mathsf{p}}\left( \Lambda _{\mathsf{p}}\right) $ depicted in Fig. \ref%
{fig:figeta}. This condition will be automatically fulfilled if to replace $%
\eta _{\mathsf{p}}\rightarrow \eta _{\mathsf{p}}\left( \Lambda _{\alpha
}\right) $ in Eq. (\ref{fia}). To this end, we introduce the new function $%
\tilde{\Phi}_{\mathsf{p}}\left( \Lambda _{\mathsf{p}}\right) \equiv \Phi _{%
\mathsf{p}}\left[ \Lambda _{\mathsf{p}},\eta _{\mathsf{p}}\left( \Lambda _{%
\mathsf{p}}\right) \right] $. Taking into account that $\Phi _{\mathsf{p}}=~1
$ for $\Lambda _{\mathsf{p}}>e/2$ one can write 
\begin{equation}
\tilde{\Phi}_{\mathsf{p}}\left( \Lambda _{\mathsf{p}}\right) =\Psi _{\mathsf{%
p}}\left( \Lambda _{\mathsf{p}}\right) +\Theta \left( \Lambda _{\mathsf{p}}-%
\frac{e}{2}\right) .  \label{FIL}
\end{equation}%
Combining Eqs. (\ref{gap0}) and (\ref{fia}) one can get the gap equation (%
\ref{gap0}) in terms of the function $\Psi _{\mathsf{p}}$ 
\begin{equation}
\ln \left( 1+\Psi _{\mathsf{p}}^{1/3}\right) -\Psi _{\mathsf{p}}^{1/3}=-\ln
\left( \Lambda _{\mathsf{p}}\right) \,\Theta \left( \Lambda _{\mathsf{p}%
}-1\right) \Theta \left( \frac{e}{2}-\Lambda _{\mathsf{p}}\right) .
\label{gamFi}
\end{equation}%
The function $\tilde{\Phi}_{\mathsf{p}}\left( \Lambda _{\mathsf{p}}\right) $
obtained as a result of numerical solution of this equation is shown in Fig. %
\ref{fig:figeta}.

For the neutron component of the superfluid mixture the integral in Eq. (\ref%
{Fi}) gives  
\begin{equation}
\Phi _{\mathsf{n}}\left( \Lambda _{\mathsf{n}},\eta _{\mathsf{n}}\right)
\,=\left( 1+\frac{3}{2}\frac{\eta _{\mathsf{n}}^{2}}{\Lambda _{\mathsf{n}%
}^{2}}\right) ^{-1}  \label{Fin}
\end{equation}%
The equation for $\tilde{\Phi}_{\mathsf{n}}\left( \Lambda _{\mathsf{n}%
}\right) \equiv \Phi _{\mathsf{n}}\left[ \Lambda _{\mathsf{n}},\eta _{%
\mathsf{n}}\left( \Lambda _{\mathsf{n}}\right) \right] $ can be found
combining Eqs. (\ref{gap2}) and (\ref{Fin}). It is convenient to convert the
unknown function to the form%
\begin{equation}
\tilde{\Phi}_{\mathsf{n}}\left( \Lambda _{\mathsf{n}}\right) =\Psi _{\mathsf{%
n}}\left( \Lambda _{\mathsf{n}}\right) +\Theta \left( \Lambda _{\mathsf{n}}-%
\sqrt{\frac{3}{2}e}\right) ,  \label{FinL}
\end{equation}%
which takes into account that $\Phi _{\mathsf{n}}=~1$ for $\Lambda _{\mathsf{%
n}}>\sqrt{\left( 3/2\right) e}$ (see appendix). Then Eq. (\ref{gap2}) can be 
written in terms of the function $\Psi _{\mathsf{n}}$%
\begin{equation}
\Psi _{\mathsf{n}}+\ln \Psi _{\mathsf{n}}=\ln \left( \frac{2}{3}\Lambda _{%
\mathsf{n}}^{2}\right) \Theta \left( \sqrt{\frac{3}{2}e}-\Lambda _{\mathsf{n}%
}\right)   \label{Psin}
\end{equation}%
The function $\tilde{\Phi}_{\mathsf{n}}\left( \Lambda _{\mathsf{n}}\right) $
obtained as a result of numerical solution of this equation is shown in Fig. %
\ref{fig:figeta2}. 

\section{Entrainment matrix}

\label{sec:em}

Let us now return to the equations for the entrainment matrix. Using of the
functions $\tilde{\Phi}_{\mathsf{p}}\left( \Lambda _{\mathsf{p}}\right) $
and $\tilde{\Phi}_{\mathsf{n}}\left( \Lambda _{\mathsf{n}}\right) $ allows
one to eliminate the gap equations from the set of equations for the
entrainment matrix thus solving only Eqs. (\ref{aa}), (\ref{ab}) but with
the replacement $\Phi _{\alpha }\left( \Lambda _{\alpha },\eta _{\alpha
}\right) \rightarrow \tilde{\Phi}_{\alpha }\left( \Lambda _{\alpha }\right) $%
. After the replacement the equations take the form%
\begin{align}
& \left( 1+\frac{1}{3}F_{1}^{\alpha \alpha }\tilde{\Phi}_{\alpha }\right)
\gamma _{\alpha \alpha }+\frac{1}{3}F_{1}^{\alpha \beta }\left( \frac{%
p_{F\beta }}{p_{F\alpha }}\right) ^{3/2}\sqrt{\frac{m_{\beta }^{\ast }}{%
m_{\alpha }^{\ast }}}\,\tilde{\Phi}_{\beta }\gamma _{\beta \alpha }  \notag
\\
& =\left( 1+\frac{1}{3}F_{1}^{\alpha \alpha }\right) \frac{m_{\alpha }}{%
m_{\alpha }^{\ast }}\,,  \label{AA}
\end{align}%
\begin{align}
& \left( 1+\frac{1}{3}F_{1}^{\alpha \alpha }\tilde{\Phi}_{\alpha }\right)
\gamma _{\alpha \beta }+\frac{1}{3}F_{1}^{\alpha \beta }\left( \frac{%
p_{F\beta }}{p_{F\alpha }}\right) ^{3/2}\sqrt{\frac{m_{\beta }^{\ast }}{%
m_{\alpha }^{\ast }}}\tilde{\Phi}_{\beta }\gamma _{\beta \beta }  \notag \\
& =\frac{1}{3}F_{1}^{\alpha \beta }\left( \frac{p_{F\beta }}{p_{F\alpha }}%
\right) ^{3/2}\frac{m_{\beta }}{\sqrt{m_{\alpha }^{\ast }m_{\beta }^{\ast }}}
\label{AB}
\end{align}%
Here $\tilde{\Phi}_{\alpha }\equiv \tilde{\Phi}_{\alpha }\left( \Lambda
_{\alpha }\right) $, and $\alpha \neq \beta $ so that if $\alpha =\mathsf{n}$
then $\beta =\mathsf{p}$ and vice-versa.

The equations (\ref{AA}), (\ref{AB}) are to be solved with taking into
account that the effective velocity $V_{\alpha }\left( \gamma _{\alpha \beta
},\mathbf{v}_{\mathsf{\alpha }s},\mathbf{v}_{\mathsf{\beta }s}\right) $ is a
complicated function of the two flow velocities, as is given in Eq. (\ref%
{absV}). Thus, the set of equations (\ref{AA}), (\ref{AB}) should be
supplemented also by the equations 
\begin{equation}
\Lambda _{\alpha }=\frac{p_{F\alpha }}{\Delta _{0}^{\left( \alpha \right) }}%
\sqrt{\gamma _{\alpha \alpha }^{2}v_{\alpha s}^{2}+2\gamma _{\alpha \alpha
}\gamma _{\alpha \beta }\mathbf{v}_{\alpha s}\mathbf{v}_{\beta s}+\gamma
_{\alpha \beta }^{2}v_{\beta s}^{2}},~\ \ \ \beta \neq \alpha ,  \label{Lv}
\end{equation}%
which reflect the fact that the gamma matrix depends actually on the
specific flow velocities $\mathbf{v}_{\mathsf{\alpha }s},\mathbf{v}_{\mathsf{%
\beta }s}$. Moreover, the gamma matrix depends on the relative direction of
the two flows.

To find a solution to Eqs. (\ref{AA}) and (\ref{AB}) we convert them to the
form

\begin{align}
\gamma _{\alpha \alpha }& ={\frac{m_{\alpha }}{m_{\alpha }^{\ast }}\frac{1}{S%
}}\,\,\left[ \left( 1+{\frac{F_{1}^{\alpha \alpha }}{3}}\right) \left( 1+{%
\frac{F_{1}^{\beta \beta }}{3}}\,\tilde{\Phi}_{\beta }\right) -\left( {\frac{%
F_{1}^{\alpha \beta }}{3}}\right) ^{2}\tilde{\Phi}_{\beta }\right] \,,
\label{s1} \\
\gamma _{\alpha \beta }& ={\frac{1}{3}}\,\,{\frac{m_{\beta }}{\sqrt{%
m_{\alpha }^{\ast }\,m_{\beta }^{\ast }}}}\,\,{\frac{1}{S}}\,\,\left( {\frac{%
p_{F_{\beta }}}{p_{F\alpha }}}\right) ^{3/2}\,\,F_{1}^{\alpha \beta }\,(1-%
\tilde{\Phi}_{\beta })\,,  \label{s2} \\
S& \equiv \left( 1+{\frac{F_{1}^{\alpha \alpha }}{3}}\,\tilde{\Phi}_{\alpha
}\right) \,\left( 1+{\frac{F_{1}^{\beta \beta }}{3}}\,\tilde{\Phi}_{\beta
}\right) -\left( {\frac{F_{1}^{\alpha \beta }}{3}}\right) ^{2}\,\tilde{\Phi}%
_{\alpha }\tilde{\Phi}_{\beta }.  \label{S}
\end{align}%
At first glance, these equations are similar to the equations obtained by
Gusakov and Haensel in \cite{gh05}, but the meaning of the functions $\tilde{%
\Phi}_{\alpha }\left( \Lambda _{\alpha }\right) $ is different. Actually,
Eqs. (\ref{s1})-(\ref{S}) recover the result obtained in \cite{gh05} only
when the flow velocities are small and $\tilde{\Phi}_{\alpha }\rightarrow 0$, see below.

%%%%%%%%%%%%%%%%%%%%%%%%%%%%%%%%%%%%%%%%%%%%%%%%%%%%%%%%%%%%%%%%%%%%

\section{Superfluid mass currents}

\label{sec:cur}

%%%%%%%%%%%%%%%%%%%%%%%%%%%%%%%%%%%%%%%%%%%%%%%%%%%%%%%%%%%%%%%%%%%%
We now turn to a calculation of the mass current caused by the superfluid
flows. Substituting Eqs. (\ref{gam}) and (\ref{nqb}) into Eq. (\ref{J}) and
performing simple integrations we find

\begin{equation}
\mathbf{j}_{\alpha }=m_{\alpha }n_{\alpha }\mathbf{V}_{\alpha }\left[ 1-%
\tilde{\Phi}_{\alpha }\left( \Lambda _{\alpha }\right) \right] ,  \label{JFi}
\end{equation}%
where the functions $\tilde{\Phi}_{\alpha }\left( \Lambda _{\alpha }\right) $
are defined in Eqs. (\ref{FIL}), and (\ref{Lv})

\begin{equation}
\mathbf{V}_{\alpha }=\sum_{\beta }\gamma _{\alpha \beta }\left( \mathbf{v}_{%
\mathsf{p}s},\mathbf{v}_{\mathsf{n}s}\right) \mathbf{v}_{\beta s},
\label{Vv}
\end{equation}%
and $\gamma _{\alpha \beta }\,\left( \mathbf{v}_{\mathsf{\alpha }s},\mathbf{v%
}_{\mathsf{\beta }s}\right) $ denotes the solution to equations (\ref{AA})-(%
\ref{Lv}) for fixed $p_{F\alpha }$ and $p_{F\beta }$. As follows from Eqs. (%
\ref{JFi}) and (\ref{Vv}) the entrainment matrix can be written in the form

\begin{equation}
{\rho }_{\alpha \beta }=\rho _{\alpha }\,\gamma _{\alpha \beta }\,\left(
\Lambda _{\mathsf{p}},\Lambda _{\mathsf{n}}\right) \left[ 1-\tilde{\Phi}%
_{\alpha }\left( \Lambda _{\alpha }\right) \right] ,  \label{rab}
\end{equation}%
where $\rho _{\alpha }=m_{\alpha }n_{\alpha }$. It represents a complicated 
nonlinear function of the velocities of superfluid flows $\mathbf{v}_{%
\mathsf{p}s}$ and $\mathbf{v}_{\mathsf{n}s}$.

It is important to notice that the right-hand side of Eqs. (\ref{s1})-(\ref%
{S}), as well as the entrainment matrix $\rho _{\alpha \beta }$ depend on $%
\Lambda _{\alpha }$ only. This greatly simplifies the problem. In practice one needs 
to solve numerically the set of equations
\begin{eqnarray*}
\Lambda _{\mathsf{p}} &=&\frac{p_{F\mathsf{p}}}{\Delta _{0}^{\left( \mathsf{p%
}\right) }}\left\vert \gamma _{\mathsf{pp}}\left( \Lambda _{\mathsf{p}%
},\Lambda _{\mathsf{n}}\right) \mathbf{v}_{\mathsf{p}s}+\gamma _{\mathsf{pn}%
}\left( \Lambda _{\mathsf{p}},\Lambda _{\mathsf{n}}\right) \mathbf{v}_{%
\mathsf{n}s}\right\vert  \\
\Lambda _{\mathsf{n}} &=&\frac{p_{F\mathsf{n}}}{\Delta _{0}^{\left( \mathsf{n%
}\right) }}\left\vert \gamma _{\mathsf{np}}\left( \Lambda _{\mathsf{p}%
},\Lambda _{\mathsf{n}}\right) \mathbf{v}_{\mathsf{p}s}+\gamma _{\mathsf{nn}%
}\left( \Lambda _{\mathsf{p}},\Lambda _{\mathsf{n}}\right) \mathbf{v}_{%
\mathsf{n}s}\right\vert 
\end{eqnarray*}%
to find $\Lambda _{\mathsf{p}}\left( \mathbf{v}_{\mathsf{p}s},\mathbf{v}_{%
\mathsf{n}s}\right) $ and $\Lambda _{\mathsf{n}}\left( \mathbf{v}_{\mathsf{p}%
s},\mathbf{v}_{\mathsf{n}s}\right) $ as functions of realistic flow
velocities $\mathbf{v}_{\mathsf{p}s}$ and $\mathbf{v}_{\mathsf{n}s}$. Than,
the entrainment matrix can be calculated with the aid of Eqs. (\ref{s1})-(\ref%
{S}) and (\ref{rab}).

%%%%%%%%%%%%%%%%%%%%%%%%%%%%%%%%%%%%%%%%%%%%%%%%%%%%%%%%%%%%%%%%%%%%%%%%%%%%%%%%%%%%

\section{Limiting cases and applications}

\label{sec:lim}

\subsection{Limiting case of small velocities}

%%%%%%%%%%%%%%%%%%%%%%%%%%%%%%%%%%%%%%%%%%%%%%%%%%%%%%%%%%%%%%%%%%%%%%%%%%%%%%%%%%%%

If the flow velocities are small as compared to the critical velocity, given
in Eq. (\ref{Vc1}) we obtain $\Lambda _{\alpha }\left( \mathbf{v}_{\mathsf{%
\alpha }s},\mathbf{v}_{\mathsf{\beta }s}\right)\ll 1$ and thus $\tilde{\Phi}%
_{\alpha }=0$. Substituting this into Eqs. (\ref{AA}), (\ref{AB}) we get%
\begin{align}
{\rho }_{\alpha \alpha }& =\rho _{\alpha }\frac{m_{\alpha }}{m_{\alpha
}^{\ast }}\left( 1+\frac{1}{3}F_{1}^{\alpha \alpha }\right) ,  \label{baa} \\
{\rho }_{\alpha \beta }& =\rho _{\beta \alpha }=\frac{p_{F\alpha
}^{2}p_{F\beta }^{2}}{9\pi ^{4}}m_{\alpha }m_{\beta }f_{1}^{\alpha \beta },
\label{bab}
\end{align}%
in agreement with the results obtained in \citet{Betal96,gh05}.

In the opposite limiting case, when $V_{\alpha }\left( \mathbf{v}_{\mathsf{%
\alpha }s},\mathbf{v}_{\mathsf{\beta }s}\right) $ tends to the
critical velocity, we obtain $\tilde{\Phi}_{\alpha }\rightarrow
1 $ which leads to ${\rho }_{\alpha \beta }=0$ because the superfluidity
disappears.

\subsection{Mixture of superfluid protons and non-superfluid neutrons}

We now turn to solving of Eqs. (\ref{AA})-(\ref{Lv}) in some particular
cases typical for the nucleon matter in superfluid cores of neutron stars.
Such a calculation requires a knowledge of the Landau parameters for a
nucleon matter in beta equilibrium. This is a separate complicated problem,
which is beyond the scope of our consideration. Since our goal is to
illustrate the importance of the nonlinear effects, we shall employ the
parameters derived in \citet{gkh09a}. Namely we consider the \textsf{npe}
matter with a density of nucleons $n_{b}=n_{\mathsf{n}}+n_{\mathsf{p}%
}=3n_{0} $, where $n_{0}=0.16$ fm$^{-3}$ is the normal nuclear density,
suggesting that the asymmetry parameter $\delta =\left( n_{\mathsf{n}}-n_{%
\mathsf{p}}\right) /n_{b}=0.837$. For this case it was obtained $F_{1}^{%
\mathsf{nn}}=-1,F_{1}^{\mathsf{pp}}=-0.55,F_{1}^{\mathsf{np}}=F_{1}^{\mathsf{%
pn}}=-0.3$.  The superfluid transition temperatures for neutrons and protons are not
reliably known. We adopt typical values used in models of neutron stars $T_{c%
\mathsf{p}}=5\times 10^{9}\,\mathsf{K}$ and $T_{c\mathsf{n}}=6\times 10^{8}\,%
\mathsf{K}$. For the energy gaps in the proton and neutron superfluids at
rest we take $\Delta _{0}^{\left( \mathsf{p}\right) }=1.764T_{c\mathsf{p}}$, 
$\Delta _{0}^{\left( \mathsf{n}\right) }=1.657T_{c\mathsf{n}}$, respectively
(see footnote$^8$).

First we consider the case when only protons in the mixture of nucleons are
superfluid and condensed into a spin-singlet state. Since the neutron
superfluidity sets on at the temperature which is well below the critical
temperature for Cooper pairing of protons, i.e., $T_{c\mathsf{n}}\ll T_{c%
\mathsf{p}}$, the limit of zero temperature can serve as a good
approximation for calculation of the superfluid mass density of protons at $%
T_{c\mathsf{n}}<T\ll T_{c\mathsf{p}}$.

For normal (non-superfluid) neutrons one should put $\tilde{\Phi}_{\mathsf{n}%
}=1$ in Eqs. (\ref{AA}) and (\ref{AB}), thus obtaining $\rho _{\mathsf{nn}}=0
$ and$\ \rho _{\mathsf{np}}=\rho _{\mathsf{pn}}={0}\,$. The remaining
component of the entrainment matrix is given by 
\begin{equation}
\rho _{\mathsf{pp}}=\rho _{\mathsf{p}}\gamma _{\mathsf{pp}}\left( 1-\tilde{%
\Phi}_{\mathsf{p}}\right) ,  \label{rpp}
\end{equation}%
where $\tilde{\Phi}_{\mathsf{p}}\left( \Lambda _{\mathsf{p}}\right) $ is
given in Eqs. (\ref{FIL}) and (\ref{gamFi}), and $\gamma _{\mathsf{pp}%
}\left( \Lambda _{\mathsf{p}}\right) $ obeys the equation%
\begin{equation}
\gamma _{\mathsf{pp}}={\frac{m_{\mathsf{p}}}{m_{\mathsf{p}}^{\ast }}}\,\,%
\frac{\left( 3+F_{1}^{\mathsf{pp}}\right) \,\left( 3+F_{1}^{\mathsf{nn}%
}\,\right) -\left( F_{1}^{\mathsf{pn}}\right) ^{2}}{\left( 3+F_{1}^{\mathsf{%
pp}}\,\tilde{\Phi}_{\mathsf{p}}\right) \,\left( 3+F_{1}^{\mathsf{nn}%
}\,\right) -\left( F_{1}^{\mathsf{pn}}\right) ^{2}\,\tilde{\Phi}_{\mathsf{p}%
}\left( \Lambda _{\mathsf{p}}\right) }\,.  \label{gpp}
\end{equation}%
The function $\Lambda _{\mathsf{p}}\left( v_{\mathsf{p}s}\right) $ is to be
found from the equation 
\begin{equation}
\Lambda _{\mathsf{p}}=\frac{p_{F\mathsf{p}}}{\Delta _{0}^{\left( \mathsf{p}%
\right) }}\left\vert \gamma _{\mathsf{pp}}\left( \Lambda _{\mathsf{p}%
}\right) v_{\mathsf{p}s}\right\vert .  \label{Lp}
\end{equation}%
\begin{figure}
\includegraphics[width=\columnwidth]{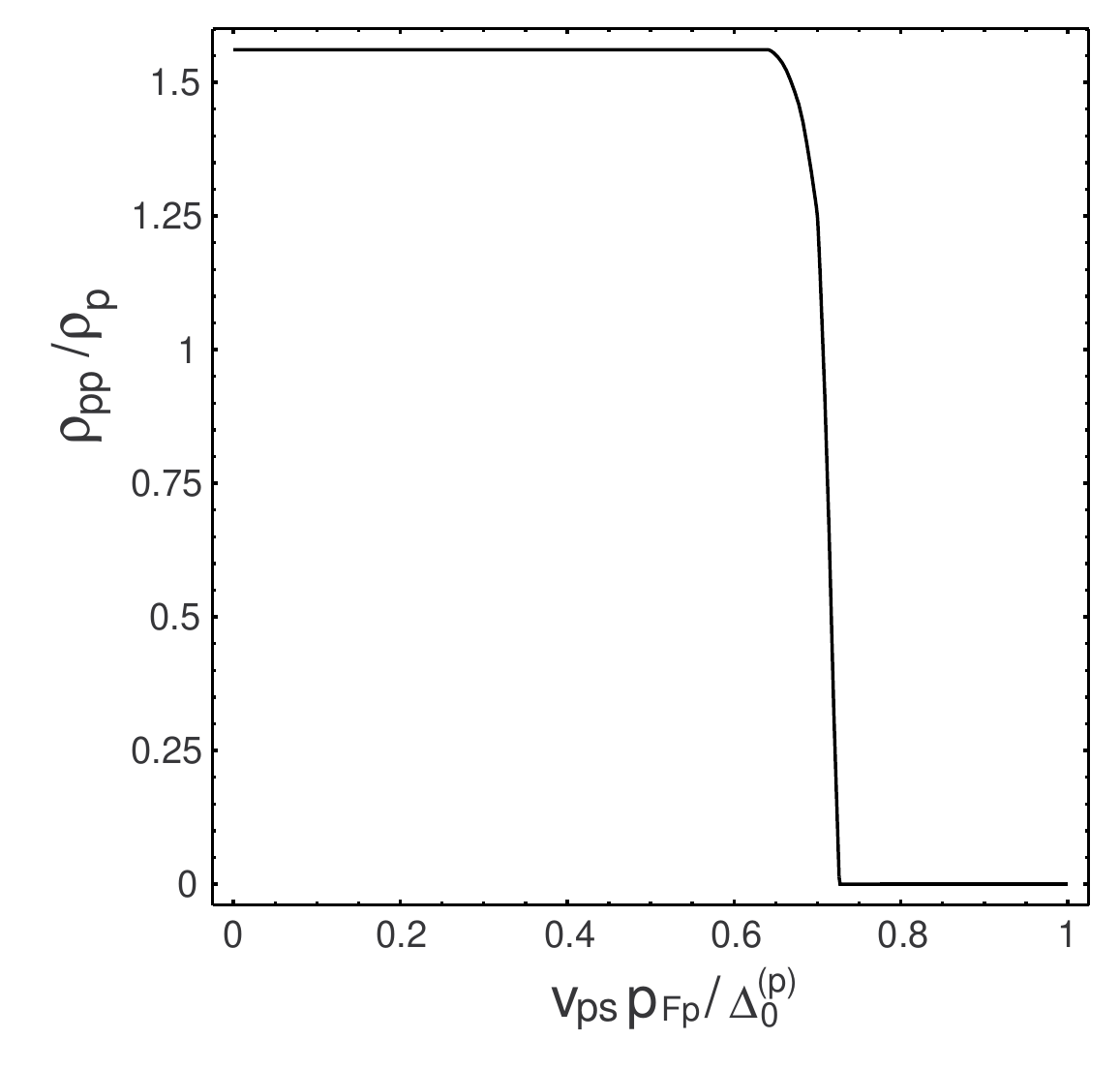}
\caption{Partial density of superfluid protons $\protect\rho _{\mathsf{pp}}/%
\protect\rho _{\mathsf{p}}$ against the superfluid velocity of protons in
the neutron-proton matter of the density $n_{b}=3n_{0}$ at zero temperature.
Only protons are assumed in a superfluid state.}
\label{fig:ropp}
\end{figure}
If we denote  
\begin{equation}
\bar{\Phi}_{\mathsf{p}}\left( v_{\mathsf{p}s}\right) =\tilde{\Phi}_{\mathsf{p%
}}\left[ \Lambda _{\mathsf{p}}\left( v_{\mathsf{p}s}\right) \right] ,
\label{Fibar}
\end{equation}%
then the partial density of superfluid protons can be written as\footnote{%
Notice, Eqs. (\ref{gpp}) and (\ref{ropp}) are similar to the equations
obtained by Gusakov and Haensel in \cite{gh05}, but the meaning of the
functions $\tilde{\Phi}_{\mathsf{p}}$ and $\bar{\Phi}_{\mathsf{p}}$ is
different.}

\begin{equation}
\frac{{\rho }_{\mathsf{pp}}}{\rho _{\mathsf{p}}}=\,{\frac{m_{\mathsf{p}}}{m_{%
\mathsf{p}}^{\ast }}}\,\,\frac{\left[ \left( 3+F_{1}^{\mathsf{pp}}\right)
\,\left( 3+F_{1}^{\mathsf{nn}}\,\right) -\left( F_{1}^{\mathsf{pn}}\right)
^{2}\right] \left( 1-\bar{\Phi}_{\mathsf{p}}\right) }{\left( 3+F_{1}^{%
\mathsf{pp}}\,\bar{\Phi}_{\mathsf{p}}\right) \,\left( 3+F_{1}^{\mathsf{nn}%
}\,\right) -\left( F_{1}^{\mathsf{pn}}\right) ^{2}\,\bar{\Phi}_{\mathsf{p}}}.
\label{ropp}
\end{equation}

The result of numerical calculation is represented in Fig. \ref{fig:ropp},
where the partial density of superfluid protons $\rho _{\mathsf{pp}}/\rho _{%
\mathsf{p}}$ is shown versus the superfluid velocity of protons. As it
follows from Fig. \ref{fig:ropp} the mass density of superfluid protons
drops very rapidly when the flow speed overcomes the value $v_{\mathsf{p}%
s}\simeq 0.65\Delta _{0}^{\left( \mathsf{p}\right) }/p_{F\mathsf{p}}$ and
completely disappears when the speed reaches $v_{\mathsf{p}s}\simeq
0.73\Delta _{0}^{\left( \mathsf{p}\right) }/p_{F\mathsf{p}}$. Notice that
this value is smaller than the known critical superfluid velocity (\ref{Vc2}%
) at which the energy gap collapses in a superfluid Fermi gas at zero
temperature.

\subsection{The case when protons and neutrons are superfluid}

It is believed that protons and hyperons in the inner core of neutron stars
condense into the isotropic $^{1}$S$_{0}$ state, while the neutron pairing
occurs in the triplet $^{3}$P$_{2}$ state with an anisotropic energy gap %
\citep{Tamagaki,Takatsuka}. 
\begin{figure*}
\centering
\includegraphics[scale=.7]{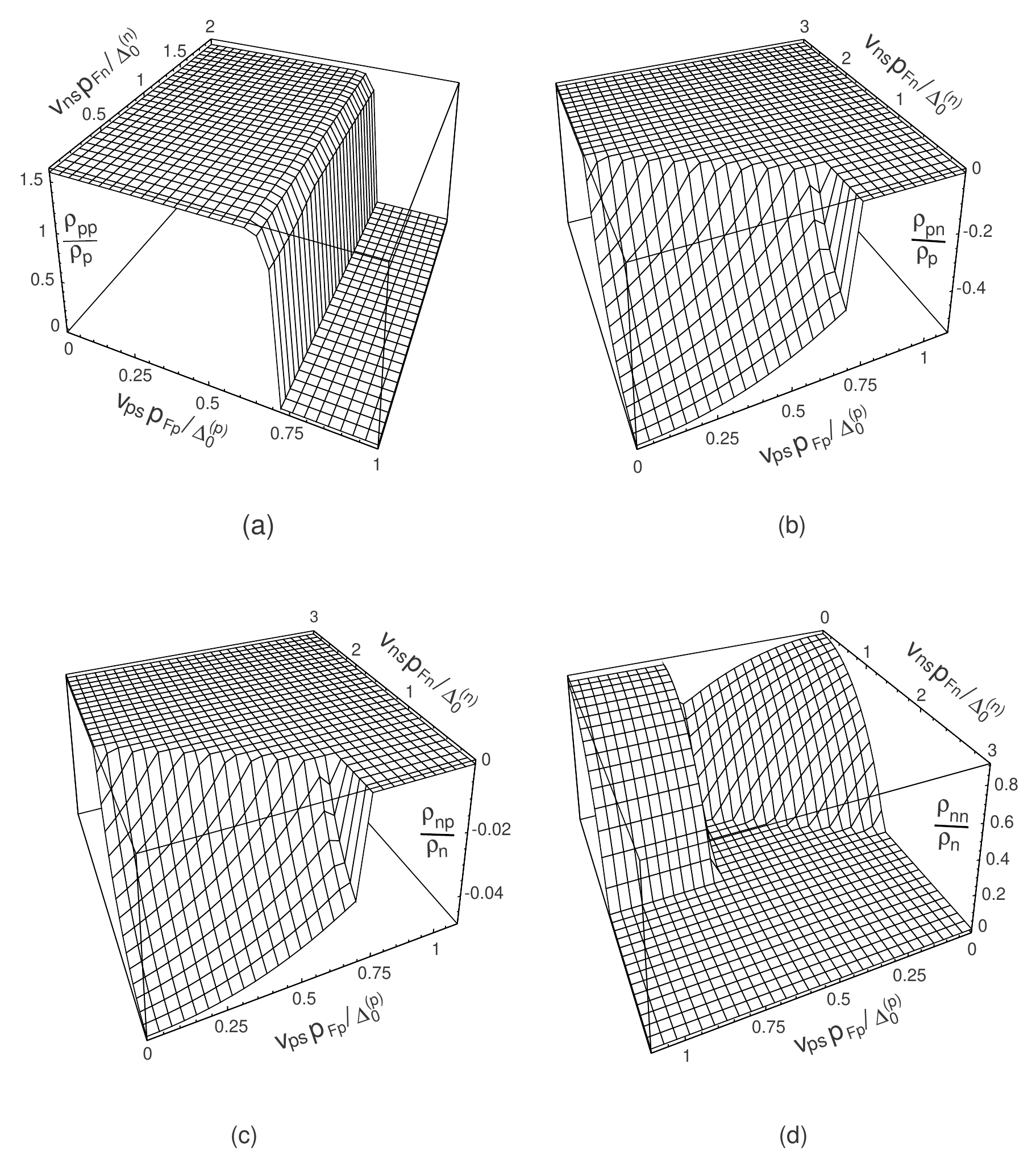}
\caption{Components of the entrainment matrix $\protect\rho_{\mathsf{pp}}$
(a), $\protect\rho_{\mathsf{pn}}$ (b), $\protect\rho_{\mathsf{np}}$ (c), $%
\protect\rho_{\mathsf{nn}}$ (d) in a mixture of superfluid protons and
neutrons at zero temperature in the case when the velocities of both
superfluid flows are parallel and have opposite directions. It is considered
the asymmetric nucleon matter of the total number density $n_b=3n_{0} $ with
the asymmetry parameter $\protect\delta =\left( n_{\mathsf{n}}-n_{\mathsf{p}%
}\right) /n_{b}=0.837$ assuming $T_{c\mathsf{p}}=5\times 10^{9}\,\mathsf{K}$ and $T_{c\mathsf{n}}=6\times 10^{8}\,\mathsf{K}$.}
\label{fig:figtotpi}
\end{figure*}
Even at rest, the $^{3}$P$_{2}$ superfluid
neutron liquid has a complicated structure, which depends on the direction
of the total angular momentum of a Cooper pair with respect to a
quantization axis $\mathbf{\hat{n}}$. 
 In a uniform system without external
fields and at absolute zero, the orientation of the quantization axis in the
superfluid at rest is arbitrary. The states of different orientation are
degenerate. For equilibrium at a non-zero temperature this leads to the
formation of a loose domain structure \citep{br60}, where each domain has a
preferred axis and these domains are randomly oriented. This fact is
normally used for the superfluids at rest in order to simplify the
calculations \citep[see{,e.g.}][]{bhy01}. 
In the work \citep{gh05} this trick is used in the
calculation of the entrainment matrix of a superfluid neutron-proton
mixture. It is assumed that on the average the entrainment matrix is
isotropic and can be calculated in the same way as in the case of the
isotropic pairing but with some effective energy gap. Such an approach can
be helpful only in the limit of small velocities of the superfluid flows,
because the velocity dependence of the energy gap in the neutron superfluid
is very different in the cases of isotropic and anisotropic pairing. This is
clearly seen from a comparison of Fig. \ref{fig:figeta} and Fig. \ref{fig:figeta2}. 
Therefore, we will consider a more realistic model in which
the pairing of neutrons in the moving superfluid liquid occurs in an
anisotropic $^{3}$P$_{2}$ state with a maximum value of the total angular
momentum $\left\vert M\right\vert =2$, and the quantization axis is directed
along the motion of the flow (see appendix).

We again consider the \textsf{npe} matter in beta equilibrium with a density
of nucleons $n_{b}=n_{\mathsf{n}}+n_{\mathsf{p}}=3n_{0}$ with the
interaction parameters, as in the previous example. The result of this
calculation is demonstrated in Fig. \ref{fig:figtotpi}, where the
entrainment matrix components are shown as functions of the two velocities
of superfluid flows of protons and neutrons. The computation was carried out
for the case when superflows of neutrons and protons move in opposite
directions.

\begin{figure*}
\centering
\includegraphics[scale=.7]{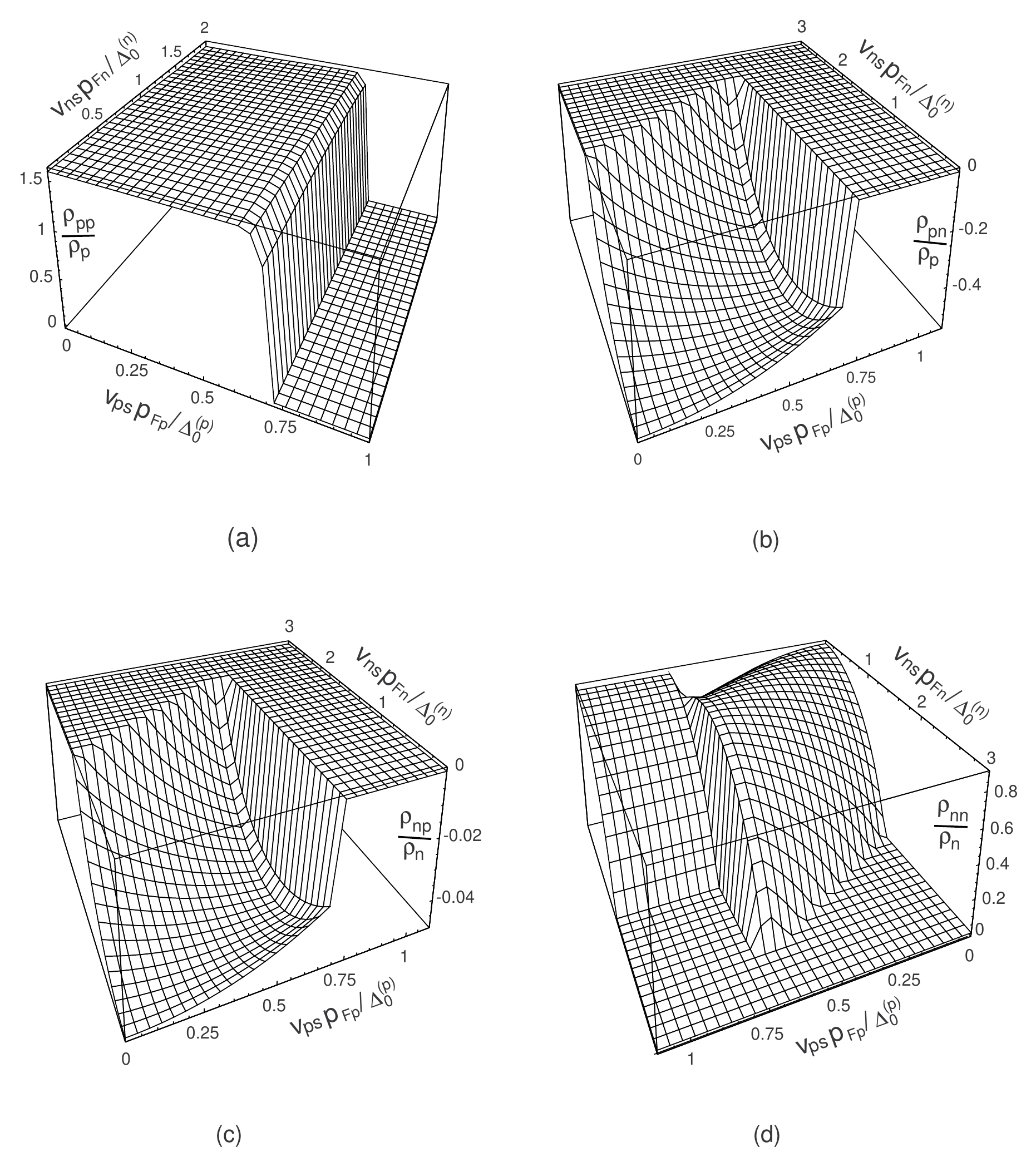}
\caption{Same as in Fig. \protect\ref{fig:figtotpi} but for the case when
superflows of protons and neutrons move in the same direction.}
\label{fig:figtot0}
\end{figure*}

Notice that the velocities of the two superfluid flows are scaled
differently in the plots. Namely, the superfluid velocity of the proton flow 
$v_{\mathsf{p}s}$ is indicated in units of $\Delta _{0}^{\left( \mathsf{p}%
\right) }/p_{F\mathsf{p}}$, while the superfluid velocity of the neutron
flow $v_{\mathsf{n}s}$ is given in units of $\Delta _{0}^{\left( \mathsf{n}%
\right) }/p_{F\mathsf{n}}$. In practice, these two scales can be very
different. For the case under consideration one can easily find that $\Delta
_{0}^{\left( \mathsf{n}\right) }/p_{F\mathsf{n}}\ll \Delta _{0}^{\left( 
\mathsf{p}\right) }/p_{F\mathsf{p}}$.

As can be seen, the $\rho _{\mathsf{pp}}$ component is almost
independent of the velocity of superfluid neutron flow but strongly depends
on the superfluid velocity of the proton flow. The $\rho _{\mathsf{pp}}$
value drops rapidly after the superfluid velocity of the proton flow exceeds
approximately $0.65\Delta _{0}^{\left( \mathsf{p}\right) }/p_{F\mathsf{p}}$
and vanishes completely when the velocity reaches the value $0.73\Delta
_{0}^{\left( \mathsf{p}\right) }/p_{F\mathsf{p}}$. Notice that this value is
substantially smaller than the known critical superfluid velocity (\ref{Vc2}%
) at which the energy gap vanishes in a superfluid Fermi gas at zero
temperature. Such a simple form of the $\rho _{\mathsf{pp}}$ component is
due to the fact that, according to Eq. (\ref{Lv}), protons give the dominant
contribution to the dimensionless effective velocity $\Lambda _{\mathsf{p}}$.

The off-diagonal mixed components of the entrainment matrix coincide to one
another, $\rho _{\mathsf{pn}}=\rho _{\mathsf{np}}$, in accordance with the
general theory \citep{ab75}. Negative values of these components vanish
rapidly when the superfluid velocity of the proton flow exceeds $0.65\Delta
_{0}^{\left( \mathsf{p}\right) }/p_{F\mathsf{p}}$ and/or the superfluid
velocity of the neutron flow exceeds $0.8\Delta _{0}^{\left( \mathsf{n}%
\right) }/p_{F\mathsf{n}}$.

The effective velocity $V_{\mathsf{n}}\left( \mathbf{v}_{\mathsf{n}s},%
\mathbf{v}_{\mathsf{p}s}\right) $ is governed by the magnitudes of both the
superfluid velocities. Therefore the component $\rho _{\mathsf{nn}}$ rapidly
drops along with growing of both the velocities. This is true, however, as
long as both nucleon constituents are in superfluid states. When the
superfluid velocity of the proton flow reaches $v_{\mathsf{p}s}\simeq
0.7\Delta _{0}^{\left( \mathsf{p}\right) }/p_{F\mathsf{p}}$ protons become
normal (non-superfluid), and their contribution to the effective velocity $%
V_{\mathsf{n}}$ is eliminated. The dependence of the entrainment matrix on
the superflow velocity becomes very similar to the case, when only one of
the components is superfluid in the mixture of nucleons. Namely, $\rho _{%
\mathsf{pn}}=\rho _{\mathsf{np}}=0$, $\rho _{\mathsf{pp}}=0$, and the $\rho
_{\mathsf{nn}}$ value becomes independent of the proton velocity $v_{\mathsf{%
p}s}$ but drops rapidly along with growing of $v_{\mathsf{n}s}$.

Let us assume now that the superflows of neutrons and protons move in the
same direction. Results of the calculation for this case are demonstrated in
Fig. \ref{fig:figtot0}, where the components of the entrainment matrix are
depicted as functions of the dimensionless velocities of the superfluid
flows of protons and neutrons. We see that the $\rho _{\mathsf{pp}}$
component of the entrainment matrix is only slightly distinct from that
component in Fig. \ref{fig:figtotpi}, while remaining components are
modified substantially due to the change of the relative direction of superfluid
flows.

The domain of velocities where the mixed components, $\rho _{\mathsf{pn}%
}=\rho _{\mathsf{np}}$, differ from zero is a little larger than in Fig. \ref%
{fig:figtotpi}. The explanation for this is the negative values of the
off-diagonal elements of the $\gamma $-matrix. In this case, for the same
directions of velocities $\mathbf{v}_{\mathsf{n}s}$ and $\mathbf{v}_{\mathsf{%
p}s}$, the effective velocity $V_{\mathsf{n}}$ is less than in the case of
their opposite directions.

The behavior of the $\rho _{\mathsf{nn}}$ component can be interpreted in
the same way as in the previous case. The velocity dependence of the
component reduces to a more simple form when the proton flow speed $v_{%
\mathsf{p}s}$ exceeds $0.73\Delta _{0}^{\left( \mathsf{p}\right) }/p_{F%
\mathsf{p}}$, and the proton superfluidity disappears. In that domain the
component $\rho _{\mathsf{nn}}$ becomes independent of the proton flow
velocity. Furthermore, $\rho _{\mathsf{pn}}=\rho _{\mathsf{np}}=0$, $\rho _{%
\mathsf{pp}}=0$, and $\rho _{\mathsf{nn}}$ drops rapidly when the superfluid
velocity of the neutron flow $v_{\mathsf{n}s}$ exceeds $1.6\Delta
_{0}^{\left( \mathsf{n}\right) }/p_{F\mathsf{n}}$.

As it follows from Figs. \ref{fig:figtotpi} and \ref{fig:figtot0}, at zero temperature, the linear approximation is accurate in a wide range of the superfluid flow velocities only for the $\rho_{\mathsf{pp}}$ component of the entrainment matrix that does not depend on the neutron flow velocity and has a plateau up to the superfluid proton velocity of about 
$0.65\Delta_{0}^{\left (\mathsf{p} \right)}/p_{F\mathsf{p}}$. Notice, this magnitude is model dependent of the values of the Landau parameters used.
For the remaining components the domain near the origin of the velocity axes, where the linear approximation can be employed is very small. These components of the entrainment matrix, can be calculated in the linear approximation only for 
$\mathbf{v}_{\mathsf{p}s}\ll \Delta_{0}^{\left(\mathsf{p}\right)}/p_{F\mathsf{p}}$ and 
$\mathbf{v}_{\mathsf{n}s}\ll\Delta_{0}^{\left(\mathsf{n}\right)}/p_{F\mathsf{n}}$.

\section{Summary and conclusion}

\label{sec:conc}

In the frame of the Fermi-liquid theory we have derived equations for the
entrainment matrix in a superfluid mixture of nucleons at arbitrary
velocities of superfluid flows. The obtained equations account for the
velocity dependence of the energy gaps and, therefore, are highly nonlinear.
The problem is solved for the case of zero temperature in the assumption of
an isotropic $^{1}$S$_{0}$ pairing of protons and anisotropic $^{3}$P$_{2}$
pairing of neutrons. We have considered the model, where the pairing of neutrons
in a moving superfluid liquid occurs in the state with a maximum value of
the total angular momentum $\left\vert M\right\vert =2$, and the
quantization axis is directed along the motion of the flow. Arguments in
favour of this model are discussed in appendix.

It is found, that in a mixture of two interacting superfluid flows, the
energy gap of each species $\alpha =\mathsf{p,n}$ depends on the two
superfluid speeds simultaneously.\footnote{%
This property was first predicted qualitatively in \citep{gk13}.} This
dependence is realized by means of the effective velocities for each species 
$V_{\mathsf{\alpha }}\left( \mathbf{v}_{\mathsf{n}s},\mathbf{v}_{\mathsf{p}%
s}\right) $ which are complicated functions of the realistic flow velocities 
$\mathbf{v}_{\mathsf{n}s},\mathbf{v}_{\mathsf{p}s}$. It is shown that the
motion of superfluid condensate leads to instability of the superfluid state
manifested in the appearance of Bogoliubov excitations at zero
temperature. For the case of $^{1}$S$_{0}$ pairing we have shown that, at
zero temperature, there exist two critical flow velocities. The first one,
which is given in Eq. (\ref{Vc1}), represents the velocity of the flow at
which the system becomes unstable. At this velocity the
minimum energy of Bogoliubov excitations vanishes. As a result such
excitations are spontaneously created in pairs and accumulate in the
superfluid system. The number of broken Cooper pairs
increases rapidly when the flow velocity builds up. All the pairs are
destroyed when the effective velocity reaches the second critical value,
which is given in Eq. (\ref{Vc2}). This is the well known critical velocity
at which the energy gap collapses in the superfluid Fermi gas. In the
considered case of triplet pairing of neutrons creation of the Bogoliubov
excitations starts immediately with the motion beginning. In both cases the
excitations result in appearance of the normal component and the energy gap
decrease.

Although the equations for the entrainment matrix are highly nonlinear they
admit a simple analytical solution in the limiting case of small velocities
of superfluid flows. In this case we get the results obtained in %
\citet{Betal96,gh05}.

For arbitrary velocities of the superfluid flows the analytical solution of
the nonlinear equations is impossible. Therefore, to demonstrate the
importance of the nonlinear approach, we have solved the equations
numerically for some particular cases typical for the nucleon matter in
superfluid cores of non-rotating  neutron stars. 
Such a calculation requires a knowledge
of the Landau parameters for a nucleon matter in beta equilibrium. This is a
separate problem, which is beyond the scope of our consideration. Since our
goal is to illustrate the importance of the nonlinear effects, we have
adopted the parameters derived in \citet{gkh09a}. Namely, we have considered
the \textsf{npe} matter with the density three times larger the normal
nuclear density, suggesting that the asymmetry parameter $\delta =0.837$.
Numerical calculation was carried out for two cases.

In the first case, it was assumed that only protons are superfluid while the
neutron liquid remains in a normal (non-superfluid) state. In this situation
the limit of zero temperature can serve as a good approximation for
calculation of the superfluid mass density of protons in the case $T_{c%
\mathsf{n}}<T\ll T_{c\mathsf{p}}$. Our computation has shown that the mass
density of superfluid protons drops very rapidly when the flow speed
overcomes the value $v_{\mathsf{p}s}\simeq 0.65\Delta _{0}^{\left( \mathsf{p}%
\right) }/p_{F\mathsf{p}}$ and completely disappears when the speed reaches $%
v_{\mathsf{p}s}\simeq 0.73\Delta _{0}^{\left( \mathsf{p}\right) }/p_{F%
\mathsf{p}}$.

Next we have considered the case when both species of nucleons are
superfluid. In this case the limit of zero temperature can be a reasonable
approximation for the entrainment matrix at temperatures much below the
critical temperature at which the neutron superfluidity sets on $T\ll T_{c%
\mathsf{n}}$. We found that the components of the entrainment matrix possess
a highly nonlinear dependence on speeds of the two superflows
simultaneously. Vanishing of the superfluidity in the superfluid mixture
takes place at the flow velocities of the order of (but not equal to) the
critical values known for single-component superfluids.

Vanishing of the superfluidity in a moving superfluid mixture can be crucial
for oscillating neutron stars. As it follows from Figs. \ref{fig:figtotpi}
and \ref{fig:figtot0} the domain near the origin of the velocity axes, where
the linear approximation can be employed is very small. On the other hand,
theoretical estimates of the neutron star pulsations \citep{kg14,gk13,gyg05} show
that even at a sufficiently small amplitude of oscillations, the relative
velocity of superfluid and normal components of the neutron matter can
exceed several times the critical value at which the superfluidity dies out.
This can affect the oscillations themselves, making them highly nonlinear.
Detailed discussions of the role of an entrainment in neutron star dynamics can
be found in \citet{pr04,cch05,cc06}.

The results obtained in the present paper can be generalized to the case of
finite temperature. This can be done making use of the temperature dependent
distribution functions (\ref{f}) in Eqs. (\ref{Fi}), (\ref{gapeq}), and (\ref%
{trgap}). The corresponding specific calculations require more numerical
work. Evidently, at finite temperatures the entrainment matrix should vanish
at smaller velocities of the superfluid flows.

%%%%%%%%%%%%%%%%5

%%%%%%%%%%%%%%%%%

\appendix

\section{$^{3}$P$_{2}$ energy gap in a superfluid flow of neutrons}

Let us consider the gap equation for a system of neutral fermions
interacting through an attractive spin-orbit potential. Let us consider a
uniform superfluid flow of neutrons of a velocity $\mathbf{v}_{s}=\mathbf{q}%
/m$ in the rest frame of the normal component. For simplicity we consider
the gap equation only for the case when the order parameter $\tilde{\Delta}%
_{\alpha \beta }\left( \mathbf{p}\right) $ is a unitary matrix in spin
space, $\left( \alpha ,\beta =\uparrow ,\downarrow \right) $, which
satisfies the condition%
\begin{equation}
\sum_{\gamma }\tilde{\Delta}_{\alpha \gamma }\left( \mathbf{p}\right) \tilde{%
\Delta}_{\gamma \delta }^{\dagger }\left( \mathbf{p}\right) =\delta _{\alpha
,\delta }\tilde{D}^{2}\left( \mathbf{p}\right) ,  \label{uc}
\end{equation}%
where $\tilde{D}^{2}(\mathbf{p})$ is real. Here and below, the tilde above a
letter indicates values that depend on the velocity of the superfluid flow.
The unitarity condition implies that the superfluid state under
consideration retains time reversal symmetry and does not have, for example,
spin polarization. For this important case the general equation for the
order parameter in a superfluid flow valid for any type of the pairing was
derived in \cite{ft72} in the BCS-Gor'kov approximation (see Eq. (12) of
that work). We consider the case of $^{3}$P$_{2}$ pairing, caused by
spin-orbit interactions. In other words we assume that the effective
interaction is most attractive for $S=1,L=1,J=2$ and that there is no tensor
coupling involving the states with different $S$ and $L$.

Since the spin-orbit interaction conserves the total momentum, its matrix
element can be written in the form%
\begin{align}
& \left\langle \mathbf{p}_{1}\alpha ,\mathbf{p}_{2}\beta \left\vert
V\right\vert \mathbf{p}_{4}\delta ,\mathbf{p}_{3}\gamma \right\rangle  
\notag \\
& =\delta _{\mathbf{p}_{1}+\mathbf{p}_{2},\mathbf{p}_{3}+\mathbf{p}%
_{4}}V_{\alpha \gamma ,\beta \delta }\left( \frac{\mathbf{p}_{1}-\mathbf{p}%
_{2}}{2}\mathbf{,}\frac{\mathbf{p}_{4}-\mathbf{p}_{3}}{2}\right) 
\label{Vme}
\end{align}

It is adopted (see e.g. \cite{Takatsuka,ft72,kcz01}) to write $V_{\alpha
\gamma ,\beta \delta }\left( \mathbf{p,p}^{\prime }\right) $ in the $^{3}$P$%
_{2}$ channel in the form of an expansion 
\begin{equation}
V_{\alpha \gamma ,\beta \delta }\left( \mathbf{p,p}^{\prime }\right) =%
{\large \upsilon }\left( p,p^{\prime }\right) \sum_{M}\Phi _{1,\alpha \beta
}^{2M}\left( \mathbf{\hat{p}}\right) \Phi _{1,\gamma \delta }^{2M\ast
}\left( \mathbf{\hat{p}}^{\prime }\right) ,  \label{VFI}
\end{equation}%
where $\Phi _{L,\alpha \beta }^{JM}\left( \mathbf{\hat{p}}\right)$~with 
$\mathbf{\hat{p}=p}/p$ are the standard spin-angle matrices 
\begin{align}
\Phi _{L,\alpha \beta }^{JM}\left( \mathbf{\hat{p}}\right) & \equiv
\sum_{M_{S}+M_{L}=M}\left( \frac{1}{2}\frac{1}{2}\alpha \beta |SM_{S}\right) 
\notag \\
& \times \left( SLM_{S}M_{L}|JM\right) Y_{L,M_{L}}\left( \mathbf{\hat{p}}%
\right) .  \label{FI}
\end{align}%
In our case it is more convenient to use vector notations. Let $\mathbf{b}%
_{M}\left( \mathbf{n}\right) $ be the vectors in the spin space that
generate the standard spin-angle matrices in accordance with%
\begin{equation}
\mathbf{b}_{M}(\mathbf{\hat{p}})\boldsymbol{\sigma }\left( i\sigma
_{y}\right) =\sqrt{8\pi }\Phi
_{1,\alpha \beta }^{2M}\left( \mathbf{\hat{p}}\right) ,  \label{saf}
\end{equation}%
where $\boldsymbol{\sigma }=\left( \sigma _{x},\sigma _{y},\sigma
_{z}\right) $ are Pauli spin matrices. There are five $^{3}$P$_{2}$ states
with a projection of the total angular momentum to the quantization axis $%
M=0,\pm 1,\pm 2$. The corresponding vectors are of the following explicit
form 
\begin{align}
\mathbf{b}_{0}\left( \mathbf{\hat{p}}\right) & =\sqrt{1/2}\left( -\hat{p}%
_{x},-\hat{p}_{y},2\hat{p}_{z}\right) ,  \notag \\
\mathbf{b}_{1}\left( \mathbf{\hat{p}}\right) & =-\sqrt{3/4}\left( \hat{p}%
_{z},i\hat{p}_{z},\hat{p}_{x}+i\hat{p}_{y}\right) ,  \notag \\
\mathbf{b}_{2}\left( \mathbf{\hat{p}}\right) & =\sqrt{3/4}\left( \hat{p}%
_{x}+i\hat{p}_{y},i\hat{p}_{x}-\hat{p}_{y},0\right) ,  \notag \\
\mathbf{b}_{-M}\left( \mathbf{\hat{p}}\right) & =\left( -\right) ^{M}\mathbf{%
b}_{M}^{\ast }\left( \mathbf{\hat{p}}\right) .  \label{bm}
\end{align}

The angular dependence of $\mathbf{b}_{M}(\mathbf{\hat{p}})$ is represented
by Cartesian components of the unit vector $\mathbf{\hat{p}=p}/p$ which
involves the polar angles on the Fermi surface, $\hat{p}_{x}=\sin \theta
\cos \varphi $,\ $\hat{p}_{y}=\sin \theta \sin \varphi $,\ $\hat{p}_{z}=\cos
\theta $. The vectors are normalized by the condition%
\begin{equation}
\int \frac{d\mathbf{\hat{p}}}{4\pi }\mathbf{b}_{M^{\prime }}^{\ast }\mathbf{b%
}_{M}=\delta _{M^{\prime }M}.  \label{lmnorm}
\end{equation}%
In the vector notations one can write the matrix element of the $^{3}$P$_{2}$
interaction as
\begin{equation}
V_{\alpha \gamma ,\beta \delta }\left( \mathbf{p,p}^{\prime }\right) =%
{\large \upsilon }\left( p,p^{\prime }\right) \sum_{M}\left[ \boldsymbol{%
\sigma }\mathbf{b}_{M}(\mathbf{\hat{p}})i\sigma _{y}\right] _{\alpha \beta }%
\left[ i\sigma _{y}\boldsymbol{\sigma }\mathbf{b}_{M}^{\ast }(\mathbf{\hat{p}%
}^{\prime })\right] _{\gamma \delta }~.  \label{Gam}
\end{equation}%
The general form of the $^{3}$P$_{2}$ order parameter is known to be of the
form \citep[see{, e.g.}][]{L10a} 
\begin{equation}
\Delta _{\alpha \beta ,\mathbf{q}}\left( \mathbf{\hat{p}}\right) =\tilde{%
\Delta}\left[ \,\mathbf{\bar{b}}\left( \mathbf{\hat{p}}\right) \boldsymbol{%
\sigma }\left( i\sigma _{y}\right) \right] _{\alpha \beta },  \label{gm}
\end{equation}%
where $\tilde{\Delta}$ is a scalar amplitude of the energy gap which depends
on the temperature and velocity of the superfluid flow, and $\mathbf{\bar{b}}%
\left( \mathbf{\hat{p}}\right) $ is some real vector in spin space which we
normalize by the condition 
\begin{equation}
\int \frac{d\mathbf{\hat{p}}}{4\pi }\mathbf{\bar{b}}^{2}\left( \mathbf{\hat{p%
}}\right) =1~.  \label{Norm}
\end{equation}%
When substituting expressions (\ref{Gam}) and (\ref{gm}) into Eq. (12) of
the work \cite{ft72}, after performing spin traces one gets the equation for
the $^{3}$P$_{2}$ energy gap in the moving flow:

\begin{align}
\tilde{\Delta}\,\left( k\right) \mathbf{\bar{b}}\left( \mathbf{\hat{k}}%
\right) & =\sum_{\mathbf{p}}{\large \upsilon }\left( k,p\right) \frac{\tilde{%
\Delta}\,\left( p\right) }{2\tilde{E}_{\mathbf{p}}}\sum_{M}\mathbf{b}_{M}(%
\mathbf{\hat{k}})\left( \mathbf{b}_{M}^{\ast }(\mathbf{\hat{p}})\mathbf{\bar{%
b}}\left( \mathbf{\hat{p}}\right) \right)   \notag \\
& \times \left( \tanh \frac{\tilde{E}_{\mathbf{p}}^{+}}{2T}+\tanh \frac{%
\tilde{E}_{\mathbf{p}}^{-}}{2T}\right),  \label{gapEq}
\end{align}%
where 
\begin{equation}
\tilde{E}_{\mathbf{p}}=\left[ \xi \left( p\right) ^{2}+\tilde{\Delta}^{2}\,%
\mathbf{\bar{b}}^{2}\left( \mathbf{\hat{p}}\right) \right] ^{1/2},
\label{ek}
\end{equation}%
\begin{equation}
\tilde{E}_{\mathbf{p}}^{\pm }=\tilde{E}_{\mathbf{p}}\pm \frac{1}{m}\mathbf{%
qp,}  \label{ekpm}
\end{equation}%
and%
\begin{equation}
\xi \left( p\right) =\frac{p^{2}}{2m}-\mu \simeq v_{F}\left( p-p_{F}\right).
\label{ksip}
\end{equation}%
Here $\mu \simeq p_{F}^{2}/\left( 2m\right) $ is the chemical potential of
degenerate particles. In writing Eq. (\ref{ek}) we have neglected the terms
quadratic in $q\ll p_{F}$. Since the volume of the system $\Omega $ drops
out of the final results we hereafter set $\Omega =1$ for brevity.

We identify the energy gap as 
\begin{equation}
\tilde{D}\left( \mathbf{p}\right) =\tilde{\Delta}\bar{b}\left( \mathbf{\hat{p%
}}\right) ,  \label{D}
\end{equation}%
where $\mathbf{\bar{b}}\left( \mathbf{\hat{p}}\right) $ is chosen to be real
in accordance to the unitarity condition (\ref{uc}), and note that the gap $%
\tilde{D}\left( \mathbf{p}\right) $ depends on the direction of the
quasiparticle momentum and, in general, has nodes. The amplitude of the
energy gap $\tilde{\Delta}$ must be real up to an arbitrary overall phase
factor $e^{-i\mathbf{qR}}$. We, therefore, may adopt that the gap amplitude $%
\tilde{\Delta}\left( p\right) $ is a real function

Making use of the orthonormality condition (\ref{lmnorm}) and the identity%
\begin{equation}
\sum_{M}\left(\int \frac{d\mathbf{\hat{p}}}{4\pi } \mathbf{\bar{b}b}_{M}\right) 
\mathbf{b}_{M}^{\ast}(\mathbf{\hat{p}})=\mathbf{\bar{b}}  \label{iden}
\end{equation}%
one can recast Eq. (\ref{gapEq}) to its final form%
\begin{equation}
\tilde{\Delta}\,\left( k\right) =\sum_{\mathbf{p}}{\large \upsilon }\left(
k,p\right) \frac{\tilde{\Delta}\,\left( p\right) \mathbf{\bar{b}}^{2}\left( 
\mathbf{\hat{p}}\right) }{2\tilde{E}_{\mathbf{p}}}\left( \tanh \frac{\tilde{E%
}_{\mathbf{p}}^{+}}{2T}+\tanh \frac{\tilde{E}_{\mathbf{p}}^{-}}{2T}\right) .
\label{GAPeq}
\end{equation}

In Eq. (\ref{GAPeq}), the pairing interaction ${\large \upsilon }\left(
k,p\right) $ depends on the absolute values of the relative particle momenta
and the integration goes over infinite momentum space. However, theory of
superfluidity needs the specific form of the pairing interaction only for a
calculation of the order parameter (the energy gap) in the superfluid at
rest. The latter completely defines all the properties of the superfluid.
Let us demonstrate this for a dependence of the energy gap on the velocity
of the superfluid flow. To this end we write Eq. (\ref{GAPeq}) in the form%
\begin{equation}
\tilde{\Delta}\,\left( k\right) =\sum_{\mathbf{p}}{\large \upsilon }\left(
k,p\right) \frac{\tilde{\Delta}\,\left( p\right) \mathbf{\bar{b}}^{2}\left( 
\mathbf{\hat{p}}\right) }{\tilde{E}_{\mathbf{p}}}\left( 1-\mathcal{F}_{%
\mathbf{p+q}}-\mathcal{F}_{-\mathbf{p+q}}\right) .  \label{gapTq}
\end{equation}%
where the functions $\mathcal{F}_{\pm \mathbf{k+q}}$ represent represent the
occupation numbers of Bogoliubov excitations in the system, which are given
by the Fermi distribution functions with a chemical potential equal to zero%
\begin{equation}
\mathcal{F}_{\pm \mathbf{k+q}}=\frac{1}{\exp \left( \tilde{E}^{\pm
}/T\right) +1}.  \label{fpm}
\end{equation}%
For a superfluid at rest (i.e. for $\mathbf{q}=0$) and temperature $T=0$ the
gap equation (\ref{gapTq}) takes the form%
\begin{equation}
\Delta _{0}\,\left( k\right) =\sum_{\mathbf{p}}{\large \upsilon }\left(
k,p\right) \frac{\Delta _{0}\,\left( p\right) \mathbf{\bar{b}}^{2}\left( 
\mathbf{\hat{p}}\right) }{E_{\mathbf{p}}},  \label{gap0n}
\end{equation}%
where%
\begin{equation}
E_{\mathbf{p}}=\sqrt{\xi _{p}^{2}+\Delta _{0}^{2}\mathbf{\bar{b}}^{2}(%
\mathbf{\hat{p}})}.  \label{Ep}
\end{equation}%
Further we multiply Eq. (\ref{gapTq}) by $\Delta _{0}\,\left( k\right) $ and
Eq. (\ref{gap0n}) by $\tilde{\Delta}\,\left( k\right) $. Combining the
obtained equations we get%
\begin{align}
& \sum_{\mathbf{p}}{\large \upsilon }\left( k,p\right) \tilde{\Delta}%
\,\left( k\right) \Delta _{0}\,\left( p\right) \mathbf{\bar{b}}^{2}\left( 
\mathbf{\hat{p}}\right) \left( \frac{1}{\tilde{E}_{\mathbf{p}}}-\frac{1}{E_{%
\mathbf{p}}}\right)   \notag \\
& =\sum_{\mathbf{p}}{\large \upsilon }\left( k,p\right) \Delta _{0}\,\left(
k\right) \tilde{\Delta}\,\left( p\right) \mathbf{\bar{b}}^{2}\left( \mathbf{%
\hat{p}}\right) \frac{1}{\tilde{E}_{\mathbf{p}}}\left( \mathcal{F}_{\mathbf{%
p+q}}+\mathcal{F}_{-\mathbf{p+q}}\right) .  \label{comb}
\end{align}%
The sums on both sides of this equation rapidly converge at a distance $\sim
\Delta _{0}$ near the Fermi surface. Indeed, the functions $\mathcal{F}_{\pm 
\mathbf{p+q}}$ on the right-hand side decrease exponentially for $\xi
_{p}\gg \Delta _{0}$. The function $(\tilde{E}_{\mathbf{p}}^{-1}-E_{\mathbf{p%
}}^{-1})$ also decreases rapidly with increase of a distance from the Fermi
surface, since for $\xi _{p}\gg \Delta _{0}$ this function can be expanded
as 
\begin{equation}
\frac{1}{\tilde{E}_{\mathbf{p}}}-\frac{1}{E_{\mathbf{p}}}\simeq \frac{%
\mathbf{\bar{b}}^{2}}{2\xi _{p}^{3}}\left( \Delta _{0}^{2}-\tilde{\Delta}%
^{2}\right) .  \label{ass}
\end{equation}%
In the narrow vicinity of the Fermi surface the smooth functions may be
replaced with their values at the Fermi surface, $\Delta _{0}\,\left(
p\right) \rightarrow \Delta _{0}\,\left( p_{F}\right) \equiv \Delta _{0}$,$~%
\tilde{\Delta}\,\left( p\right) \rightarrow \tilde{\Delta}\,\left(
p_{F}\right) \equiv \tilde{\Delta}$, ${\large \upsilon }\left( k,p\right)
\rightarrow {\large \upsilon }\left( p_{F},p_{F}\right) $, and moved out
from under the summation signs. This yields

\begin{equation}
\sum_{\mathbf{p}}\mathbf{\bar{b}}^{2}\left( \mathbf{\hat{p}}\right) \left( 
\frac{1}{\tilde{E}_{\mathbf{p}}}-\frac{1}{E_{\mathbf{p}}}\right) =\sum_{%
\mathbf{p}}\mathbf{\bar{b}}^{2}\left( \mathbf{\hat{p}}\right) \frac{1}{%
\tilde{E}_{\mathbf{p}}}\left( \mathcal{F}_{\mathbf{p+q}}+\mathcal{F}_{-%
\mathbf{p+q}}\right) .  \label{eg}
\end{equation}%
The summation on the left-hand side can be done with the aid of the formula%
\begin{equation}
\int \frac{d^{3}p}{\left( 2\pi \right) ^{3}}\cdot \cdot \cdot =\frac{%
p_{F}m^{\ast }}{2\pi ^{2}}\int \frac{d\mathbf{\hat{p}}}{4\pi }\int_{-\infty
}^{\infty }d\xi _{p}\cdot \cdot \cdot .  \label{1}
\end{equation}%
This gives%
\begin{equation}
\frac{p_{F}m}{\pi ^{2}}\ln \frac{\Delta _{0}}{\tilde{\Delta}}=\sum_{\mathbf{p%
}}\frac{\mathbf{\bar{b}}^{2}\left( \mathbf{\hat{p}}\right) }{\tilde{E}_{%
\mathbf{p}}}\left( \mathcal{F}_{\mathbf{p+q}}+\mathcal{F}_{-\mathbf{p+q}%
}\right) ,  \label{trip}
\end{equation}%
where $\Delta _{0}$ is uniquely related to the temperature of the superfluid
transition $T_{c}$, which is assumed to be known, and $\tilde{\Delta}$ is
the energy gap, which for a fixed density depends on the temperature and
velocity of the superfluid flow. The distribution functions for the
Bogoliubov excitations are given by 
\begin{equation}
\mathcal{F}_{\pm \mathbf{p+q}}=\frac{1}{1+\exp \left[ \frac{1}{T}\left( 
\tilde{E}_{\mathbf{p}}\pm p_{F}\mathbf{v}_{s}\mathbf{\hat{p}}\right) \right] 
}.  \label{dis}
\end{equation}%
with%
\begin{equation}
\tilde{E}_{\mathbf{p}}=\sqrt{\xi _{p}^{2}+\tilde{\Delta}^{2}\,\mathbf{\bar{b}%
}^{2}\left( \mathbf{\hat{p}}\right) }.  \label{Et}
\end{equation}%
The solution to Eq. (\ref{trip}) allows one to find the gap $\tilde{\Delta}$
as a function of the superfluid velocity $\mathbf{v}_{s}$ at temperatures
below the critical value $T_{c}$ for the superfluidity onset.

The vector $\mathbf{\bar{b}}$ can be generally written as $\bar{b}_{i}=A_{ij}%
\hat{p}_{j}$. In the case of a unitary $^{3}$P$_{2}$ condensate the $3\times
3$ matrix $A_{ij}$ must be a real symmetric traceless tensor. It may be
specified by giving the orientation of its principal axes and its two
independent diagonal elements in its principal-axis coordinate system.

Since the ground state should be invariant under time reversal the states
with magnetic quantum numbers $\pm M$ must be populated with equal
likelihood. On the other hand, the gap tensor $A_{ij}$ should be diagonal.
This second requirement excludes the possibility of populating states with $%
M=\pm 1$. \ Therefore within the preferred coordinate system, there exist
two simple solutions of Eq. (\ref{GAPeq}).

For the first solution, which represents a condensation of the pairs into
the state with $M=0$, we have $\mathbf{\bar{b}}(\mathbf{\hat{p}})=\mathbf{b}%
_{0}(\mathbf{\hat{p}})$ and $\mathbf{\bar{b}}^{2}(\mathbf{\hat{p}}%
)=1/2\left( 1+3\cos ^{2}\theta \right) $. The second solution of Eq. (\ref%
{GAPeq}) corresponds to $\left\vert M\right\vert =2$. In this case $\mathbf{%
\bar{b}}(\mathbf{\hat{p}})=1/\sqrt{2}\left[ \mathbf{b}_{2}(\mathbf{\hat{p}})+%
\mathbf{b}_{-2}(\mathbf{\hat{p}})\right] $, and $\mathbf{\bar{b}}^{2}(%
\mathbf{\hat{p}})=3/2\left( 1-\cos ^{2}\theta \right) $. As is well known
the two solutions, with $M=0$ and $\left\vert M\right\vert =2$, are almost
degenerate in the neutron superfluid at rest.

We now turn to study these states in the moving condensate. In the
superfluid at rest all directions for the principal-axis of the gap matrix
are equivalent because the corresponding states are degenerate. However, the
superfluid motion removes the degeneration. The terms proportional to 
$\mathbf{pv}_{s}$, in the quasiparticle energy (\ref{ekpm}), make the direction of the
quantization axis along the motion more favourable. Therefore one may
consider that at zero temperature the preferred direction for the
principal-axis of the gap matrix is specified by the direction of superflow
motion, i.e. $Oz\parallel \mathbf{q}$. This choice seems reasonable because
the direction of the flow is the only preferred direction which exists in
the uniform system in the absence of external fields.

Now we need to choose between the states $M=0$ and $\left\vert M\right\vert
=2$. To this end, we can compare the maximum values of the superfluid flow
velocity $\mathbf{v}_{s\max }=\mathbf{q}_{\max }/m$ at which the energy gap
vanishes in each of the two cases. To find it we consider Eq. (\ref{trip})
for $\tilde{\Delta}/\Delta _{0}\ll 1$. In the limit $T\rightarrow 0$,~this
equation becomes of the form%
\begin{equation}
\ln \frac{\Delta _{0}^{2}}{\tilde{\Delta}^{2}}=2\int_{0}^{1}d\cos
\theta \,\int_{0}^{\infty }du\frac{\mathbf{\bar{b}}^{2}\Theta \left( q_{\max
}\xi _{0}\cos \theta -u\right) }{\sqrt{u^{2}+\left( \tilde{\Delta}%
^{2}/\Delta _{0}^{2}\right) \mathbf{\bar{b}}^{2}}},  \label{tau0}
\end{equation}%
where $\xi _{0}=v_{F}/\Delta _{0}$ is the coherence length of correlated
fermions at zero temperature, and $\Theta \left( z\right) $ is a unit-step
function, $\Theta \left( z\right) =1$ if $z\geq 0$ and $\Theta \left(
z\right) =0$ otherwise. Since in the rest frame the states $M=0$ and $%
\left\vert M\right\vert =2$ are almost degenerate one can put $\xi
_{0}^{\left( M_{J}=0\right) }\simeq \xi _{0}^{\left( \left\vert
M_{J}\right\vert =2\right) }=\xi _{0}$.

Doing the last integral in the limit $\tilde{\Delta}/\Delta _{0}\rightarrow
0~$ we get the relation 
\begin{equation}
\int_{0}^{1}\,\mathbf{\bar{b}}^{2}\ln \frac{4q_{\max }^{2}\xi _{0}^{2}\cos
^{2}\theta }{\mathbf{\bar{b}}^{2}}d\cos \theta =0.  \label{qeq}
\end{equation}

For a $^{3}$P$_{2}$ condensation with $M_{J}=0$ one has $\mathbf{\bar{b}}%
^{2}=\left( 1/2\right) \left( 1+3\cos ^{2}\theta \right) $. In this case,
from Eq. (\ref{qeq}) we find limiting value of $q_{\max }^{\left(
M_{J}=0\right) }$ at which the energy gap vanishes at zero temperature 
\begin{equation}
q_{\max }^{\left( M_{J}=0\right) }\xi _{0}=\frac{\sqrt{2}}{2}e^{\frac{1}{27}%
\pi \sqrt{3}+\frac{1}{6}}=1.\,\allowbreak 021\,9\,.  \label{qm0}
\end{equation}%
For a condensation with $\left\vert M_{J}\right\vert =2$ one has $\mathbf{%
\bar{b}}^{2}=\left( 3/2\right) \sin ^{2}\theta $. In this case Eq. (\ref{qeq}%
) gives%
\begin{equation}
q_{\max }^{\left( \left\vert M_{J}\right\vert =2\right) }\xi _{0}^{\left(
\left\vert M_{J}\right\vert =2\right) }=\sqrt{\frac{3}{2}e}=2.\,\allowbreak
019\,.  \label{qm2}
\end{equation}%
A comparison of Eqs. (\ref{qm0}) and (\ref{qm2}) allows one to prefer the mode, 
where the moving $^{3}$P$_{2}$ condensate exists in the state with 
$\left\vert M_{J}\right\vert =2$. It is obviously true at least at 
$q>q_{\max }^{\left(M_{J}=0\right) }$.

%%%%%%%%%%%%%%%%%%55

\end{document}